%% file: main.tex
\documentclass[10pt,journal,compsoc]{IEEEtran}
\usepackage{algorithmic}
\usepackage[linesnumbered,ruled,vlined]{algorithm2e}

\usepackage{array}
\usepackage[caption=false,font=normalsize,labelfont=sf,textfont=sf]{subfig}
\usepackage{textcomp}
\usepackage{stfloats}
\usepackage{url}
\usepackage{verbatim}
\usepackage{graphicx}
\usepackage{multirow}
\usepackage{colortbl}
\usepackage{float}
\usepackage{booktabs}
\usepackage{mathtools}
\usepackage[utf8]{inputenc}
\usepackage{framed}
\usepackage{tikz}
\usepackage{enumitem}
\usepackage{xcolor}
\usepackage{soul}
\usepackage{amsmath,amssymb,amsfonts,amsthm}

\def\BibTeX{{\rm B\kern-.05em{\sc i\kern-.025em b}\kern-.08em
    T\kern-.1667em\lower.7ex\hbox{E}\kern-.125emX}}
\usepackage{balance}

\newtheorem{theorem}{Theorem}
\newtheorem{lemma}{Lemma}
\definecolor{cGreen}{RGB}{0,150,0}%{128,255,128}
\definecolor{brown}{RGB}{139,64,0}
\newcommand{\bs}[1]{\boldsymbol{#1}}
\newcommand{\gray}[1]{{\color{gray}~#1 }}

\newcommand{\ourname}{Condensing {Pre-augmented} Recommendation Data via Lightweight Policy Gradient Estimation}

\newcommand{\ournameAbbr}{{DConRec}}

\begin{document}

\title{\ourname}
\author{Jiahao Wu, Wenqi Fan, Jingfan Chen, Shengcai Liu, Qijiong Liu, \\Rui He, Qing Li, \IEEEmembership{Fellow,~IEEE}, and Ke Tang, \IEEEmembership{Fellow,~IEEE}
\IEEEcompsocitemizethanks{\IEEEcompsocthanksitem J. Wu, R. He, S. Liu. and K. Tang are with the Guangdong Provincial Key
Laboratory of Brain-Inspired Intelligent Computation and Department of
Computer Science and Engineering, Southern University of Science and
Technology, Shenzhen 518055, China. \protect\\ %(Corresponding author: Ke Tang.)
E-mail: notyour\_mason@outlook.com, her2018@mail.sustech.edu.cn, liusc3@sustech.edu.cn, and tangk3@sustech.edu.cn
\IEEEcompsocthanksitem J. Wu, W. Fan, J. Chen, Q. Liu and Q. Li are with the Department of Computing, The Hong Kong Polytechnic University, Hong Kong.\protect\\
E-mail: notyour\_mason@outlook.com, wenqi.fan@polyu.edu.hk, jingfan.c-\\hen@connect.polyu.hk, liu@qijiong.work, and csqli@comp.polyu.edu.hk
% \IEEEcompsocthanksitem  is with the Centre for Frontier AI Research (CFAR), Agency for Science, Technology and Research (A*STAR), Singapore 138632.\protect\\
% E-mail:
}% <-this % stops a space
\thanks{(Corresponding author: Shengcai Liu.)
%Recommended for acceptance by xxx \protect\\
%Digital Object Identifier no. xxx
}}

% \markboth{Journal of \LaTeX\ Class Files,~Vol.~18, No.~9, xxxx~2024}%
\markboth{IEEE TRANSACTIONS ON KNOWLEDGE AND DATA ENGINEERING}
{How to Use the IEEEtran \LaTeX \ Templates}

\IEEEtitleabstractindextext{%
\begin{abstract}
Training recommendation models on large datasets requires significant time and resources. 
It is desired to construct concise yet informative datasets for efficient training. 
Recent advances in dataset condensation show promise in addressing this problem by synthesizing small datasets. However, applying existing methods of dataset condensation to recommendation has limitations: (1) they fail to generate discrete user-item interactions, and (2) they could not preserve users' potential preferences. To address the limitations, we propose a lightweight condensation framework tailored for recommendation (\textbf{DConRec}), focusing on condensing user-item historical interaction sets. Specifically, we model the discrete user-item interactions via a probabilistic approach and design a pre-augmentation module to incorporate the potential preferences of users into the condensed datasets. While the substantial size of datasets leads to costly optimization, we propose a lightweight policy gradient estimation to accelerate the data synthesis. Experimental results on multiple real-world datasets have demonstrated the effectiveness and efficiency of our framework. Besides, we provide a theoretical analysis of the provable convergence of DConRec. The implementation is available at: \url{https://github.com/JiahaoWuGit/DConRec}.
%The code will be released upon publication.
\end{abstract}
}

\maketitle
\IEEEdisplaynontitleabstractindextext
% \IEEEdisplaynontitleabstractindextext has no effect when using
% compsoc under a non-conference mode.

% For peer review papers, you can put extra information on the cover
% page as needed:
% \ifCLASSOPTIONpeerreview
% \begin{center} \bfseries EDICS Category: 3-BBND \end{center}
% \fi
%
% For peerreview papers, this IEEEtran command inserts a page break and
% creates the second title. It will be ignored for other modes.
\IEEEpeerreviewmaketitle

% % \clearpage
% \section{Introduction}
% \IEEEPARstart{W}{elcome} \cite{wu-et-al:DcRec_cikm22}

\input{sections/intro}
\input{sections/preliminaries}
\input{sections/methodology}
\input{sections/experiments}

\input{sections/relatedWork}

% \balance
\input{sections/conclusion}

\input{sections/acknowledge}
% \input{sections/Appendix}

% \balance
\bibliographystyle{IEEEtran}
\bibliography{references}
% \newpage
\input{sections/author-bio}
\clearpage
\input{sections/Appendix}
% \balance
\end{document}

%% file: sections/intro.tex
\section{Introduction}
Recommender systems~\cite{he-et-al:NeuMF17he,tkde2022Diffnet++,fan2019graphRec,tkde2023lianghaoXia,tkde2024sslRec} play a pivotal role in alleviating information overload predicament in the era of data explosion. Trained on the large-scale user-item historical interaction sets~\cite{fan2020graphTKDE,sigir2019NGCF,wenqi23LLMRec,TKDE2023lewuSurvey} , these
models have demonstrated the ability to provide personalized recommendations across various websites (e.g., Amazon, Taobao). Despite their promising performance, training models on large datasets is computationally expensive. 
Such training cost even becomes prohibitive when repeated training is required. 
To avoid the high computational cost, it is desired to train models on small substitute datasets. 
\input{sections/sub-sections/experiments/fig-intro-preAugANDRuntime}

Dataset condensation (also denoted as dataset distillation) sheds light on a direction for mitigating the aforementioned issue. 
The aim of dataset condensation is to synthesize a small dataset while preserving the essential knowledge for model training~\cite{zhaoICLR2021DC,doscond-kdd2022,tongzhouw2018datasetdistillation,nips2021InfiniteDD}. This enables models to achieve comparable performance to those trained on the full dataset.
Particularly, one of the most representative methods, DC~\cite{zhaoICLR2021DC}, synthesizes the dataset by gradient matching, which aligns the gradients of network parameters trained on small synthetic and large real data. 
Motivated by DC~\cite{zhaoICLR2021DC}, a line of works have been proposed in various domains (e.g., images~\cite{icml2022DCSSL,icml2022DCViaEffi,icml2021DC-Siamese} and graph~\cite{doscond-kdd2022,wang2023braveTpami,iclr2022gcond}).
It has been demonstrated that {such methods} can significantly reduce dataset size with minimal performance drop.
Despite its effectiveness {in these domains}, dataset condensation for recommendation has been rarely investigated.

To fill this gap, we investigate dataset condensation for recommendation, focusing on condensing interaction set. 
However, directly applying dataset condensation for recommendation faces challenges:
(1) {Non-differentiable discrete data:} Most existing works are designed for continuous data (\textit{e.g.}, images) or data with continuous features (\textit{e.g.}, graph node features), but hardly focus on discrete binary data (\textit{e.g.}, interaction data in recommendation). Condensing discrete interactions with those frameworks could pose a non-differentiable problem.
(2) {Limited users'  preference preservation:} 
Existing works generate synthetic data based on the original dataset. However, in recommendation, \textit{directly condensing} original user-item interaction set may be inadequate to preserve users' potential preference, since users are exposed to a limited number of items in the original dataset.
Such a method may produce low-quality datasets, as shown in Figure~\ref{fig:intro-data-qualities} {(e.g., the quality of condensed Ciao from \textit{w/o PA} declines 46$\%$, compared to \textit{PA}).} 

To address the aforementioned challenges, we propose a novel dataset condensation framework (named \textbf{\ournameAbbr}) for recommendations, which is established on a bi-level optimization pipeline. {The bi-level optimization involves dual loops:} the inner loop optimizes the recommendation model, while the outer loop condenses the discrete interaction data. 

To enable the discrete outer optimization differentiable, we probabilistically reparameterize the user-item interactions. Furthermore, to incorporate users' potential preference into condensation, we introduce a pre-augmentation module. 

While this framework has shown promise in tackling the limitations of existing methods, it has encountered a new challenge. Specifically, the bi-level formulation introduces computational inefficiency, since the inner optimization involves multiple full training and the outer optimization involves verbose gradient flow (Figure~\ref{fig:intro-dc-compare}), which is computationally expensive. This inefficiency is especially pronounced with a large number of user-item interactions. For instance, the prevalent solver for this bi-level optimization, gradient matching (GradMatch), exhibits low efficiency in condensation for recommendation, as shown in Figure~\ref{fig:intro-condensing-efficiencies}. 

To tackle this challenge, we propose a new solver, named lightweight policy gradient estimation (LPGE). The LPGE enhances the condensation efficiency in two ways: (1) it adopts a lightweight update scheme for the  inner model training, reducing inner loop iterations, and (2) it utilizes a policy gradient estimator (PGE) for the data updates, reducing the computational cost during outer loop (Figure~\ref{fig:intro-dc-compare}).

We empirically verify the effectiveness and efficiency of DConRec via extensive experiments and theoretically examine its provable convergence. We summarize main contributions as follows:
\begin{itemize}[leftmargin=*]
    \item Investigate a novel problem of condensing user-item interactions for recommendation, where we adopt probabilistic reparameterization to continualize the discrete bi-level optimization problem.
    \item Propose a lightweight dataset condensation framework (DConRec) for recommendation. To be specific, we propose a pre-augmentation module to preserve users' potential preference. Additionally, we propose a lightweight policy gradient estimator for the condensation to avoid computational inefficiency. Further, we examine the provable convergence property of DConRec.
    \item Demonstrate the effectiveness and efficiency of our framework in condensing recommendation datasets, reducing the dataset size by 75\% and approximating 98\% of the original performance on Dianping and about 90\% on other datasets. Besides, our framework is significantly faster than the prevalent gradient matching solver (e.g., $8\times$ in dataset Ciao for updating the data for 100 epochs).
\end{itemize}

% The rest of the paper is structured as follows: Section~\ref{sec:notations-preliminary} introduces the notations and preliminaries of recommendation and dataset condensation. Section~\ref{sec:framework} presents the proposed framework. Section~\ref{sec:experiments} shows the experimental results. Related works are reviewed in section~\ref{sec:related-work}. Finally, we conclude the paper in section~\ref{sec:conclusion}.

% \red{you could totally follow the logic line of dosCond, i.e., in the followings: potential solution(introduction of DD/DC) -> key challenges of DD/DC in recommendation -> you proposed methods and your framework -> what you have done and a brief overview of the total paper.}
\input{sections/sub-sections/fig-intro-dc-compare}

%% file: sections/sub-sections/experiments/fig-intro-preAugANDRuntime.tex
\begin{figure*}[t]
\vskip -0.15in
\centering

\subfloat[Pre-aug.'s Impact]{\label{fig:intro-data-qualities}\includegraphics[width=0.56\columnwidth]%[width=0.67\columnwidth]
{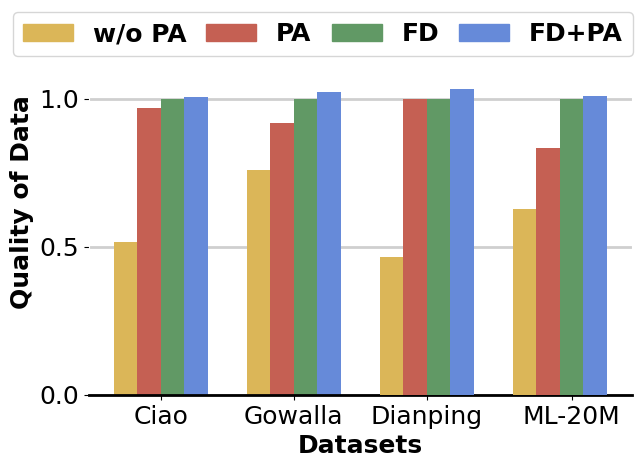}}
\hspace{0.015\columnwidth}
\subfloat[Efficiency]{\label{fig:intro-condensing-efficiencies}\includegraphics[width=0.5\columnwidth]%[width=0.62\columnwidth]
{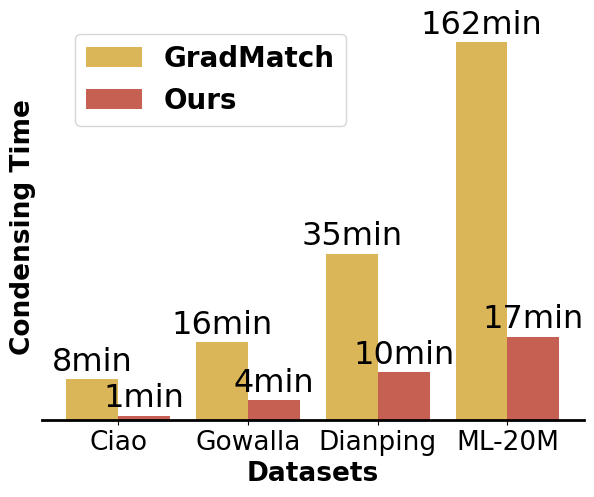}}%{figures/experiments/intro/intro-run-time-1.png}}
\hspace{0.015\columnwidth}
\subfloat[Effectiveness]{\label{fig:intro-condensing-effectiveness}\includegraphics[width=0.5\columnwidth]%[width=0.7\columnwidth]
{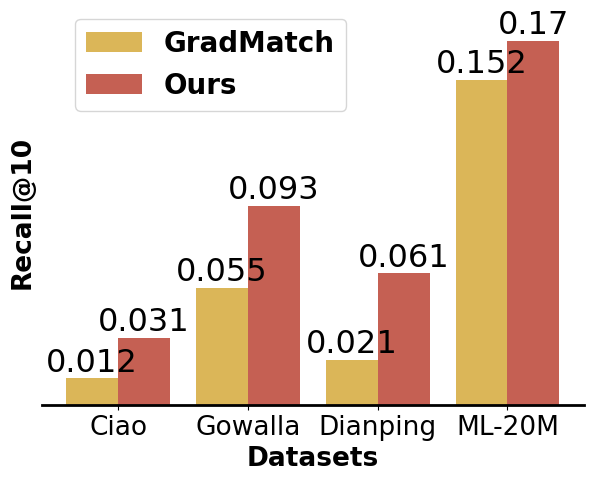}}
% \hspace{0.004\columnwidth}

\vskip -0.15in
\caption{(i) Pre-augmentation significantly enhances the quality of the condensed datasets (\textnormal{subfig.a}). The quality is evaluated by comparing the model's performance trained on the datasets to the performance with full data. {\textbf{w/o PA}}: dataset condensed from original dataset, {\textbf{PA}}: dataset condensed from pre-augmented dataset, {\textbf{FD}}: original dataset, {\textbf{FD+PA}}: pre-augmented dataset. (ii) Our method yields better efficiency (\textbf{subfig.b}) and effectiveness (\textbf{subfig.c}) than the prevalent gradient matching method.}%in terms of dataset condensation

\label{fig:intro-preAugANDRuntime}
% \label{fig:intro-condensing-efficiencies}
\vskip -0.15in
\end{figure*}
% \textcolor[RGB]{219,182,88}{\textbf{w/o PA}}
% \textcolor[RGB]{197,96,83}{\textbf{PA}}
% \textcolor[RGB]{97,153,101}{\textbf{FD}}
% \textcolor[RGB]{102,138,217}{\textbf{FD+PA}}

% \begin{figure}[t]
% % \vskip -0.1in
% \centering

% \subfloat[Efficiency]{\label{fig:intro-condensing-efficiencies}\includegraphics[width=0.333\columnwidth]{figures/experiments/intro/intro-run-time-kuang.png}}
% % \hspace{0.0025\columnwidth}
% \subfloat[Effectiveness]{\label{fig:intro-condensing-effectiveness}\includegraphics[width=0.333\columnwidth]{figures/experiments/intro/intro-performance-kuang.png}}
% % \hspace{0.004\columnwidth}
% \subfloat[Pre-aug.'s Impact]{\label{fig:intro-data-qualities}\includegraphics[width=0.333\columnwidth]{figures/experiments/intro/intro-ablation-kuang.png}}

% \vskip -0.05in
% \caption{(i) Our method yields superior efficiency (\textnormal{subfig.a}) and effectiveness (\textnormal{subfig.b}) compared to the prevalent gradient matching method in terms of dataset condensation. (ii) Pre-augmentation significantly enhances the quality of the condensed datasets (\textnormal{subfig.c}). The quality is evaluated by comparing the model's performance trained on the datasets to the performance with full data. \textcolor[RGB]{219,182,88}{w/o PA}: dataset condensed from original dataset, \textcolor[RGB]{197,96,83}{PA}: dataset condensed from pre-augmented dataset, \textcolor[RGB]{102,138,217}{FD+PA}: pre-augmented dataset.} 

% \label{fig:intro-preAugANDRuntime}
% % \label{fig:intro-condensing-efficiencies}
% \vskip -0.2in
% \end{figure}

%% file: sections/sub-sections/fig-intro-dc-compare.tex
\begin{figure}[]
\centering
\vskip -0.1in
% {\includegraphics[width=1.0\linewidth]{{figures/intro-dd-compar2.pdf}}}
{\includegraphics[width=0.95\linewidth]{{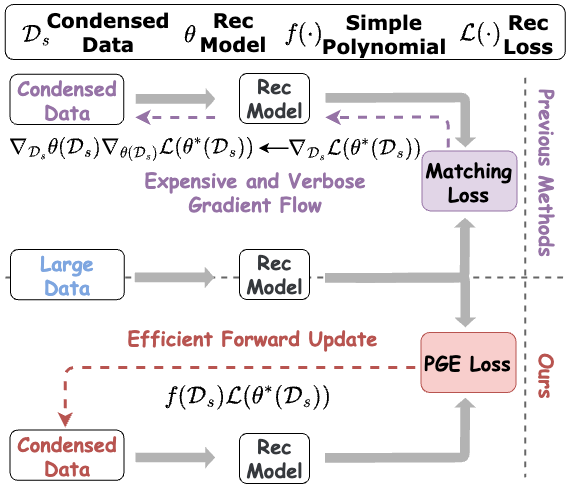}}}
\vskip -0.1in
\caption{{\textbf{Expensive Backward Update}} \textit{vs}  {\textbf{Efficient Forward Update}}: the backward update used in previous methods involves expensive and verbose gradient flow while our proposed framework adopts efficient forward update, inherently avoiding the high computational expenses.}
% \red{1. pre-augmentation: show the technical details 2. re-draw the figure of large dataset and data pool back to the user-item type 3.PGE written in math form. You need to highlight the constrast}\blue{use pen to draw a draft figure first} \red{add illustration to $u_1$ and $i_m$}}%\red{haven't explicitly express the efficiency, maybe should highlight the efficiency and the differentiation}}
\vskip -0.2in
\label{fig:intro-dc-compare}
\end{figure}

% \textcolor[RGB]{150,115,166}{\textbf{Expensive Backward Update}}
% \textcolor[RGB]{184,84,80}{\textbf{Efficient Forward Update}}

%% file: sections/preliminaries.tex
\section{Notations and Preliminaries}
\label{sec:notations-preliminary}
\textbf{Notations.}
% \red{you need to carefully re-check the notations in other sections}
Let $\mathcal{D} = \{(u,i)|u\in \mathcal{U}, i\in \mathcal{I}, y_{u,i}=1\}$ denote the observed user-item interactions, where $\mathcal{U}$ is the set of users, $\mathcal{I}$ is the set of items, and $y_{u,i}={1}$ indicates that user $u$ has interacted with item $i$. $f_{{\theta}}$ denotes the recommender model with ${\theta}$ being the trainable parameters. We denote the synthesized set of user-item interactions after condensation by $\mathcal{D}_s= \{(u,i)|u\in \mathcal{U}, i\in \mathcal{I}, m_{u,i}=1\}$, which is parameterized by $\mathbf{M}$ and $\mathbf{M}\in \{1,0\}^{|\mathcal{U}|\times|\mathcal{I}|}$ is a binary matrix. In this matrix, $m_{u,i}=1$ indicating that in the synthesized set user $u$ has interacted with item $i$. The interactions discovered by pre-augmentation form a pseudo dataset $\mathcal{D}_{ps}$, representing users' potential preference. The data pool $\mathcal{D}_{p}$ used for condensation consists of the pseudo dataset $\mathcal{D}_{ps}$ and the original dataset $\mathcal{D}$. 
% The final condensed dataset will be compensated with interactions of long-tailed users $\mathcal{D}_{t}$: $\mathcal{D}'_{s} = \mathcal{D}_s\cup\mathcal{D}_{t}$. 
The condensation ratio is $r=\frac{|\mathcal{D}_s|}{|\mathcal{D}|}$.\\
% and $|\mathcal{D}_{t}|\ll |\mathcal{D}_s|$.\\%, we have $r\approx \frac{|\mathcal{D}_{cond}|}{|\mathcal{D}|}$. %For simplicity, we denote the condensed dataset with the ratio of $r$ by $\mathcal{D}_s^r$. 
\textbf{Preliminaries.} 
{\textit{Recommendation.}}
% \red{refer to Efficient Bi-Level Optimization for Recommendation Denoising}
% \red{the prliminaries about the recommendation is poor}
Given a user-item interaction set $\mathcal{D}$, the goal of recommendation is to model the preference of users and then recommend items that users may be interested in. Aiming to learn optimal model parameter $\theta^*$, the training objective of recommendation model $f$ based on dataset $\mathcal{D}$ is formulated as:
\begin{align}
    \label{eq:recommendation-formulation}
    {\theta}^* = \underset{{\theta}}{\arg \min\; } \mathcal{L}(f_{{\theta}}(\mathcal{D})),
\end{align}
where $\mathcal{L}(\cdot)$ is the recommendation loss 
(i.e. BPR loss~\cite{rendle-et-al:bpr2009}).

{\textit{Dataset Condensation.}} The goal of dataset condensation~\cite{zhaoICLR2021DC,icml2022DCSSL,doscond-kdd2022,New2023GraphCondense,ctrDD2023RecSys} is to condense a large dataset into a small synthetic dataset, such that model trained on the small dataset can achieve comparable performance to a model trained on the large full dataset~\cite{ndss2023backdoorDD}. In this paper, we define the goal of dataset condensation for recommendation as synthesizing a small interaction set $\mathcal{D}_s=\{y_{u,i}|u\in \mathcal{U}, i\in \mathcal{I}, m_{u,i}=1\}$, such that a model trained on $\mathcal{D}_s$ can achieve close performance to the one trained on the full dataset $\mathcal{D}$, which is much larger than $\mathcal{D}_s$. Therefore, the objective of condensing recommendation data is formulated as follows:
\begin{align}
    \label{eq:bi-level-rec-data-condense}
    \min_{\mathcal{D}_s} \mathcal{L}(f_{{\theta}_s}(\mathcal{D})), \text { s.t. } {\theta}_s=\underset{{\theta}}{\arg\min\;}\mathcal{L}(f_{{\theta}}(\mathcal{D}_s)),
\end{align}
where ${\theta}_s$ denotes the model parameters trained on the synthesized dataset $\mathcal{D}_s$ and $\mathcal{L}(\cdot)$ is the recommendation loss.%\red{introduce gradient matching}
% refer to DATASET CONDENSATION WITH GRADIENT MATCHING, ICLR2021

{\textit{Gradient Matching for Dataset Condensation.}} Gradient matching~\cite{zhaoICLR2021DC,ctrDD2023RecSys,doscond-kdd2022} is one of the most popular scheme for dataset condensation. Concretely, it tries to reduce the difference of model gradients w.r.t. large-real data and small-synthetic data for model parameters at every training epoch. Let $\theta_t$ denotes the model parameters at $t$-epoch. The condensation objective is formulated as: 
\begin{align}    
\begin{gathered}
\min _{\mathcal{D}_s} \underset{\theta_0 \sim P_{\theta_0}}{\mathbb{E}}\left[\sum_{t=0}^{T-1} D\left(\nabla_\theta \mathcal{L}\left(f_{\theta_t}(\mathcal{D}_s)\right), \nabla_\theta \mathcal{L}\left(f_{\theta_t}(\mathcal{D})\right)\right)\right] \\
\text { s.t. } \theta_{t+1}=\operatorname{opt}_\theta\left(\theta_t, \mathcal{D}_s\right),
\end{gathered} 
\end{align}%\ell
where $D(\cdot,\cdot)$ is a distance function, $T$ is the number of model training steps, $\operatorname{opt}_\theta(\cdot)$ is the optimization operator, $\mathcal{L}(\cdot)$ is the loss for dowstream tasks (e.g., BPR loss~\cite{rendle-et-al:bpr2009} in recommendation tasks) and $P_{\theta_0}$ is the distribution of $\theta_0$. The efficient one-step gradient matching scheme~\cite{doscond-kdd2022,ctrDD2023RecSys,KDD23bilevelDenoise} compared in this paper sets $T=1$.%, enabling generalization of $\mathcal{D}_s$ cross various initializations.

% \clearpage

%% file: sections/methodology.tex
\section{Methodologies}
\label{sec:framework}
% \input{sections/sub-sections/merged-framework}
% In this section, we elaborate on our proposed \ourname, which is illustrated in Figure~\ref{fig:condense-frameowork}. First, we present the module of generating data pool, summarized in Figure~\ref{fig:pseudo-data-generation}. Then, the condensation framework for recommendation is presented. Thereafter, the optimization for this framework is introduced.  %\red{you could write a subsection about overview of the proposed framework: referring to the paper Self-Supervised Dual-Channel Attentive Network for Session-based Social Recommendation}
\subsection{Overiew of the Proposed Framework}
In this section, we present the details of our proposed DConRec framework, with overall architecture depicted in Figure~\ref{fig:pdc-framework-combine}.
In section~\ref{subsec:data pool generation}, we introduce pre-augmentation on original dataset to generate data pool, aiming to preserve the potential users' preference in the process of condensation. In section~\ref{subsec:condense-rec-data}, we introduce the probabilistic modeling of interactions, addressing the discreteness,  and the bi-level formulation of the condensation for the interaction data. 
In section~\ref{subsec:optim-os-pge}, to enhance the condensation efficiency, we propose a lightweight policy gradient estimator (LPGE) to solve the bi-level optimization and provide theoretical examination on the convergence property of DConRec. %In subsection~\ref{subsec:discussion}, we discuss the advantages of DConRec in condensing interaction data.

% Specifically, we utilize a well-trained proxy model $f_{proxy}$ to identify items that users may like. Those items and users form a pseudo dataset $\mathcal{D}_{ps}$ and the data pool $\mathcal{D}_{p}$ consist of the pseudo dataset $\mathcal{D}_{ps}$ and the original dataset $\mathcal{D}$. Then, the synthetic dataset $\mathcal{D}'_{s}$ is generated (subsection~\ref{subsec:condense-rec-data}).
% the condensation generates the synthetic dataset $\mathcal{D}'_{s}$ which consists of the synthesized set $\mathcal{D}_s$ of informative user-item interactions identified from the data pool $\mathcal{D}_{p}$ and the set $\mathcal{D}_{t}$ of compensated interactions for long-tailed users (subsection~\ref{subsec:condense-rec-data}).
% Finally, the lightweight optimization strategy is presented in subsection~\ref{subsec:optim-os-pge}.%Precisely, we assign each interaction in data pool $\mathcal{D}_{p}$

% Since the interactions are discrete, we utilize the probabilistic reparameterization to make the optimization process for condensation differential. %\red{In each iteration of the condensation, we first synthesize the condensed dataset based on the }
% As shown in Figure~xxx, the framework consists of xxx parts: first part (in subsection xxx), second part (in subsection xxx), .....
\subsection{Pre-augmenting to Generate Data Pool}
\label{subsec:data pool generation}
% \red{you revise the paper till here.}
% \input{sections/sub-sections/fig-pseudo-data}
To preserve users' potential preference in the condensed datasets, we propose to pre-augment the original dataset to generate the data pool before condensation. %Specifically,  
The data pool comprises two parts: the original interaction dataset $\mathcal{D}$ and the pseudo dataset $\mathcal{D}_{ps}$. The pseudo dataset consists of interactions between users and unexposed items identified by the proxy model $f_{p}$ as potential interests for users. We denote those items as \textit{pseudo items}. 

As shown in Figure~\ref{fig:pdc-framework-combine}a, the proxy model $f_{p}$ is first trained based on the original dataset $\mathcal{D}$. Then, the well-trained proxy model is utilized to identify pseudo items for each user $u$. Specifically, for each user $u$, the pseudo interaction set $\text{TopK}(u)$ is composed of interactions between them with their pseudo items. We formalize the pseudo interaction set for user $u$ as following:
% \red{here we may re-write the IsTopK}:
\begin{align}
    \label{eq:topk-pseudo-each-user}
    \text{TopK}(u) = \{i|i=\underset{i'}{\arg \text{IsTopK}}\;f_p(u,i^{\prime}),\;i'\in\mathcal{I}^{\prime}(u)\},
\end{align}
where $f_p(u,i')$ is the preference score predicted by proxy model. %: $score_u(i)=f_{proxy}(\mathbf{z}_u)^Tf_{proxy}(\mathbf{z}_i)$. 
$\mathcal{I}'(u)$ is the set of items that have not interacted with user $u$ in $\mathcal{D}$. Then, the pseudo dataset $\mathcal{D}_{ps}$ is:
\begin{align}
    \label{eq:pseudo-dataset}
    \mathcal{D}_{ps}=\{(u,i)|u\in \mathcal{U},i\in TopK(u)\}.
\end{align}
Then, the data pool $\mathcal{D}_{p}$ is constructed by $\mathcal{D}$ and $\mathcal{D}_{ps}$:
\begin{align}
    \label{eq:data-pool}
    \mathcal{D}_{p}=\mathcal{D}\cup\mathcal{D}_{ps}.
\end{align}
The generation of data pool is illustrated in Figure~\ref{fig:pdc-framework-combine}a.
% $f_{{proxy}}(\mathbf{z}_u)$
% The generation of a data pool is furthered by the utilization of a computationally efficient proxy recommender model, which enables the identification of unexposed items anticipated to be favored by users, consequently resulting in the formation of a pseudo dataset. We denote the pseudo dataset by $\mathcal{D}_{ps}= \{y_{ui'}|u\in \mathcal{U}, i'\in \mathcal{I}, y_{ui'}\notin \mathcal{D}\}$.
\input{sections/sub-sections/fig-datapoolAndOSPGE}
\subsection{Condensing Recommendation Dataset}
\label{subsec:condense-rec-data}

% \red{corresponding to the subsection of Bilevel Framework for Coreset Selection in ICML22 p-c-s}
\subsubsection{\textbf{Formalization}}
% \red{you revise paper till here.}
Given the data pool $\mathcal{D}_{p}$, we parameterize condensed dataset $\mathcal{D}_s$ by the generation matrix $\mathbf{M}$ and follow Eq.~\ref{eq:bi-level-rec-data-condense} to formulate the condensation into a discrete bi-level optimization:
\begin{align}
    \label{eq:bilevel_discrete}
    % \begin{gathered}
          \min _{\mathbf{M} \in \tilde{\mathcal{C}}} \tilde{\Phi}(\mathbf{M})= \mathcal{L}\left({\theta}^*(\mathbf{M})\right), 
         \text { s.t. } {\theta}^*(\mathbf{M}) = \underset{{\theta}}{\arg \min\; } \hat{\mathcal{L}}({\theta} ; \mathbf{M}),
    % \end{gathered}
\end{align}
where the inner optimization is to train model while the outer optimization is to update condensed dataset:
\begin{align}
    % % \begin{aligned}
    % \begin{gathered}
        \label{eq:bilevel_discrete_loss}
        &\mathcal{L}\left({\theta}^*(\mathbf{M})\right)=\frac{1}{|\mathcal{D}|} \sum_{{y}_{u,i}\in \mathcal{D}} \ell\left(f\left(u,i ; {\theta}^*(\mathbf{M})\right), {y}_{u,i}\right), \\
        % &\hat{\mathcal{L}}({\theta} ; \mathbf{M})=\frac{1}{|\mathcal{U}|} \sum_{{y}_{u,i}\in \mathcal{D}_{p}} \frac{1}{K_u(p)}m_{u,i} \ell\left(f\left(u,i ; {\theta}\right), {y}_{u,i}\right), \\
        &\hat{\mathcal{L}}({\theta} ; \mathbf{M})=\frac{1}{r|\mathcal{D}|} \sum_{{y}_{u,i}\in \mathcal{D}_{s}} \ell\left(f\left(u,i ; {\theta}\right), {y}_{u,i}\right), \\
         % &\mathcal{D}_{p}=\mathcal{D}\cup\mathcal{D}_{ps},
        &\tilde{\mathcal{C}}=\{\mathbf{M}|m_{u,i}=0\;or\;1,\;\lVert\mathbf{M}\rVert_0\leq\;r|\mathcal{D}|\}%,~~~%\nonumber\\
        % &
        % \mathcal{D}'_{s}=\mathcal{D}_s\cup\mathcal{D}_{t},
    % \end{gathered}
    % \end{aligned}
\end{align}
where $\tilde{\mathcal{C}}$ is the feasible region of the generation matrix $\mathbf{M}$ for synthesized dataset $\mathcal{D}_s$. 
In this paper, we set the recommendation loss to BPR loss~\cite{rendle-et-al:bpr2009}. Since we generate the synthesized data points based on $\mathcal{D}_{p}$, we permanently set $m_{u,i}=0$ while $(u,i)\notin \mathcal{D}_{p}$.
In the inner optimization, the recommender model is trained on the synthesized dataset $\mathcal{D}_s$, which is parameterized by $\mathbf{M}$; the optimal model parameter is denoted by ${\theta}^*(\mathbf{M})$. 
 In the outer optimization, the goal is to evaluate the performance of ${\theta}^*(\mathbf{M})$ on the original dataset $\mathcal{D}$ and minimize the loss on $\mathcal{D}$ to optimize $\mathbf{M}$.  %xxxxxxxxxx xxxxxxxxxx xxxxxxxxxx xxxxxxxxxx %xxxxxxxxxx xxxxxxxxxx xxxxxxxxxx xxxxxxxxxx xxxxxxxxxx xxxxxxxxxxxxxxxxxxxx xxxxxxxxxx xxxxxxxxxx xxxxxxxxxxxxxxxxxxxx
% \red{we proofread till here.}
% \input{sections/sub-sections/fig-os-pge}
\subsubsection{\textbf{Continualization}} Since it is intractable to optimize the aforementioned discrete objective via gradient, we propose to transfer the discrete optimization into a continuous one via probabilistic reparameterization~\cite{doscond-kdd2022,PCoreset-icml2022}. Specifically, we reparameterize $m_{u,i}$ as a Bernoulli random variable with probability $s_{u,i}$ to be $1$ and probability $1-s_{u,i}$ to be $0$. Therefore, we have $m_{u,i}\sim\operatorname{Bern}(s_{u,i})$ and $s_{u,i}\in [0,1]$. We assume that the preference of each user $u$ towards different items is independent, such that the variables $m_{u,i}$ are independent. In this way, we can obtain the distribution of $\mathbf{M}$: 
% \begin{align}
% p(\mathbf{M}|\mathbf{S})=\underset{u\in \mathcal{U}}{\prod} \left[\underset{i\in \mathcal{I}}{\prod} \left[(s_{u,i})^{m_{u,i}}(1-s_{u,i})^{(1-m_{u,i})}\right]\right],
% \end{align}
\begin{align}
p(\mathbf{M}|\mathbf{S})=\underset{u\in \mathcal{U}}{\prod}\underset{i\in \mathcal{I}}{\prod} (s_{u,i})^{m_{u,i}}(1-s_{u,i})^{(1-m_{u,i})},
\end{align}
where $\mathbf{S}$ is the probabilistic matrix. Since those interactions $(u,i)\notin \mathcal{D}_{p}$ will not be generated in the condensed dataset, we set $m_{u,i}=0$ for them and rewrite the above distribution:
\begin{align}
p(\mathbf{M}|\mathbf{S})=\underset{(u,i)\in \mathcal{D}_{p}}{\prod} (s_{u,i})^{m_{u,i}}(1-s_{u,i})^{(1-m_{u,i})}.
\end{align}
Therefore, we could obtain the restriction on $\mathbf{S}$ from the restriction on the values of $\mathbf{M}$: 
\begin{align}
    \mathbb{E}_{\mathbf{M}\sim p(\mathbf{M}|\mathbf{S})}\lVert\mathbf{M}\rVert_0=\underset{(u,i)\in\mathcal{D}_{p}}{\sum}s_{u,i}.
\end{align}
Based on this, we can obtain the feasible region $\mathcal{C}$ for $\mathbf{S}$. Here, with the condensation ratio $r=\frac{|\mathcal{D}_s|}{|\mathcal{D}|}$, we have:
\begin{align}
    \mathcal{C}=\{\mathbf{S}|0\leq s_{u,i}\leq 1, \lVert\mathbf{S}\lVert_1\leq r|\mathcal{D}|\}.
\end{align}
Then, the discrete optimization described in Eq.~\ref{eq:bilevel_discrete} could be re-formulated into the following continuous one:
\begin{align}
\label{eq:continuous_bilevel}
    % \begin{gathered}
        \min _{\mathbf{S} \in \mathcal{C}} \Phi(\mathbf{S})= \mathbb{E}_{p} \mathcal{L}\left({\theta}^*(\mathbf{M})\right), 
        \text { s.t. } {\theta}^*(\mathbf{M}) = \underset{{\theta}}{\arg \min\; } \hat{\mathcal{L}}({\theta} ; \mathbf{M}),
    % \end{gathered}
\end{align}
where $p$ denotes the distribution $\mathbf{M}\sim p(\mathbf{M}|\mathbf{S})$. After we get the optimal probabilistic matrix $\mathbf{S}^*$ for the interactions, we can sample the corresponding generation matrix $\mathbf{M}^*$. Then, we can generate the condensed dataset $\mathcal{D}_s$ based on $\mathbf{M}^*$. %The generation of the compensation dataset $\mathcal{D}_{t}$ for long-tailed dataset is presented in the subsequent subsection. 

% \subsubsection{\textbf{Compensation for Long-tailed Users}} To avoid amplifying the long-tailed problem, we propose to compensate the synthesized dataset with interactions between long-tailed users and their interacted items in $\mathcal{D}_{p}$. Those interactions forms the compensation dataset $\mathcal{D}_{t}$ for the long-tailed users, which could be formalized as:
% % \begin{align}
% %     \label{eq:compensation-dataset-tail}
%     $\mathcal{D}_{t}=\underset{|\mathcal{D}_u|<\epsilon}{\cup}\mathcal{D}_u,$
% % \end{align}
% where $\mathcal{D}_u$ is the interaction set of user $u$ and it is a subset of data pool $\mathcal{D}_{p}$. Besides, $\epsilon$ is the threshold of the number of interactions, utilized to identify long-tailed users.

\subsection{Optimization via Lightweight Policy Gradient Estimation}
\label{subsec:optim-os-pge}
As we can observe from Eq.~\ref{eq:continuous_bilevel}, the optimization is computationally expensive. It requires to differentiate through the model optimal, thus involving verbose gradient flow, to calculate the gradients for data updates. To be specific, according to chain-rule, the gradient calculation for the probability matrix $\mathbf{S}$ is presented in the form of:
\begin{align}
    \label{eq:chain-rule-bilevel}
    \nabla_{\mathbf{S}} \Phi(\mathbf{S}) = \nabla_{\mathbf{S}} {\theta}^*(\mathbf{M}) \nabla_{{\theta}} \mathcal{L}\left({\theta}^*(\mathbf{M})\right),
\end{align}
where $\nabla_{\mathbf{S}} {\theta}^*(\mathbf{M})$ can lead to expensive and verbose gradient flow. It needs to unroll the verbose gradient flows over multiple inner optimization steps. Thus, the computational and memory demands associated with such calculation are substantial. Moreover, multiple epochs of data updates further exacerbates the cost since it requires multiple inner model convergences.

% the gradient calculation for data updates requires multiple model convergences, which further exacerbates the cost.

To address the aforementioned two issues, we propose Lightweight Policy Gradient Estimation (LPGE). The idea is two-fold: (1) In order to avoid the onerous gradient calculation to update $\mathbf{S}$, we propose to utilize policy gradient estimator~\cite{PCoreset-icml2022}. This estimator only involves forward propagation to calculate the gradients for the probability matrix $\mathbf{S}$ updates. (2) To mitigate the computation cost from multiple inner convergences, we propose a lightweight substitute for the inner optimization, which adopts the one-step update strategy~\cite{doscond-kdd2022,KDD23bilevelDenoise}. The intuition behind is that the loss decreases after first step optimization of the inner model training is strong enough to provide comprehensive information to guide the outer data update. This technique is also verified to be effective empirically and theoretically in previous studies~\cite{doscond-kdd2022,KDD23bilevelDenoise,oneStepCoreset2021NIPS}. The  overall optimization process is summarized in algorithm~\ref{algo:training}. 

In the rest of this subsection, we elaborate on the details of policy gradient estimator and lightweight optimization for inner model training as well as providing theoretical understanding on the convergence property of DConRec.
\subsubsection{\textbf{Policy Gradient Estimator (PGE)}}To avoid the computationally expensive gradient flow of Eq.~\ref{eq:chain-rule-bilevel}, we first transform the objective term in Eq.~\ref{eq:continuous_bilevel} as following:
\allowdisplaybreaks[4]
% \begin{equation}
    \begin{align}
%     \begin{aligned}
    % \end{aligned}
    \label{eq:PGE}
    \nabla_{\mathbf{S}} \Phi(\mathbf{S}) & =\nabla_{\mathbf{S}} \mathbb{E}_{p(\mathbf{M}|\mathbf{S})} \mathcal{L}\left({\theta}^*(\mathbf{M})\right) \notag\\ \notag
    & =\nabla_{\mathbf{S}} \int \mathcal{L}\left({\theta}^*(\mathbf{M})\right) p(\mathbf{M}|\mathbf{S}) d \mathbf{M} \\ \notag
    & =\int \mathcal{L}\left({\theta}^*(\mathbf{M})\right) \frac{\nabla_{\mathbf{S}} p(\mathbf{M}|\mathbf{S})}{p(\mathbf{M}|\mathbf{S})} p(\mathbf{M}|\mathbf{S}) d \mathbf{M} \\ \notag
    & =\int \mathcal{L}\left({\theta}^*(\mathbf{M})\right) \nabla_{\mathbf{S}} \ln p(\mathbf{M}|\mathbf{S}) p(\mathbf{M}|\mathbf{S}) d \mathbf{M} \\ 
    & =\mathbb{E}_{p(\mathbf{M}|\mathbf{S})} \mathcal{L}\left({\theta}^*(\mathbf{M})\right) \nabla_{\mathbf{S}} \ln p(\mathbf{M}|\mathbf{S}),
    \end{align}
%     \end{aligned}
% \end{equation}
where $\mathcal{L}\left({\theta}^*(\mathbf{M})\right) \nabla_{\mathbf{S}} \ln p(\mathbf{M}|\mathbf{S})$ is called policy gradient. Given the inner optimum ${\theta}^*(\mathbf{M})$, i.e., the well-trained model, we can directly calculate the gradient and update $\mathbf{S}$ as following:
\begin{align}
\label{eq:gradient-probability-s}
    \mathbf{S} \leftarrow \mathcal{P}_{\mathcal{C}}\left(\mathbf{S}-\eta \mathcal{L}\left({\theta}^*(\mathbf{M})\right) \nabla_{\mathbf{S}} \ln p(\mathbf{M}|\mathbf{S})\right),
\end{align}
where $\mathcal{P}_{\mathcal{C}}(\cdot)$ is to project the updated values of probability matrix $\mathbf{S}$ back to the feasible region $\mathcal{C}$ of $\mathbf{S}$ and it is shown in algorithm~\ref{algo:projection}. 

\textbf{{Analysis on the effectiveness of PGE}}: From the design of PGE, shown in Eq.\ref{eq:chain-rule-bilevel} and Eq.\ref{eq:PGE}, we can observe that its optimization direction is towards minimizing the loss of downstream task $\mathcal{L}(\cdot)$ (e.g., BPR loss here) by synthesizing data, which is in line with the model training for good performance. 

\textbf{Analysis on the computational efficiency of PGE}: from Eq.~\ref{eq:gradient-probability-s}, we observe updating $\mathbf{S}$ involves: $\mathcal{L}\left({\theta}^*(\mathbf{M})\right)$, $\nabla_{\mathbf{S}} \ln p(\mathbf{M}|\mathbf{S})$ and $\mathcal{P}_{\mathcal{C}}(\cdot)$. Computing $\mathcal{L}\left({\theta}^*(\mathbf{M})\right)$ is efficient, only requiring forward propagation. This avoids the expensive and verbose gradient flow. Besides, the form of $\nabla_{\mathbf{S}} \ln p(\mathbf{M}|\mathbf{S})$ is quite simple and the projection $\mathcal{P}_{\mathcal{C}}(\cdot)$ is an efficient transformation as summarized in algorithm~\ref{algo:projection}. Therefore, PGE is significantly faster than previous methods for data updates.%\red{the proof for the algorithm could be written in.}

\subsubsection{\textbf{Lightweight Optimization}}
\label{subsubsec:lightweight}
To further improve the overall computational efficiency, we propose the lightweight optimization. Specifically, we adopt the one-step update strategy~\cite{doscond-kdd2022,KDD23bilevelDenoise} for the inner model update and the policy gradient estimation in the Eq.~\ref{eq:PGE} could be re-written as following:
\begin{align}
        \label{eq:one-step-PGE}
    \nabla_{\mathbf{S}} \Phi(\mathbf{S}) =\mathbb{E}_{p(\mathbf{M}|\mathbf{S})} \mathcal{L}\left({\theta}_1(\mathbf{M})\right) \nabla_{\mathbf{S}} \ln p(\mathbf{M}|\mathbf{S}),
\end{align}
where ${\theta}_1\left(\mathbf{M}\right)$ is obtained via one-step inner model update: 
\begin{align*}
    {\theta}_1\left(\mathbf{M}\right)\leftarrow {\theta}_0\left(\mathbf{M}\right)-\nabla_{{\theta}_0}{\theta}_0\left(\mathbf{M}\right). 
\end{align*}
Therefore, the gradient calculation for the probability matrix $\mathbf{S}$ (Eq.~\ref{eq:gradient-probability-s}) could be re-formalized as following:
\begin{equation}
    \label{eq:one-step-gradient-s}
    \mathbf{S} \leftarrow \mathcal{P}_{\mathcal{C}}\left(\mathbf{S}-\eta \mathcal{L}\left({\theta}_1(\mathbf{M})\right) \nabla_{\mathbf{S}} \ln p(\mathbf{M}|\mathbf{S})\right).
\end{equation}
\subsubsection{\textbf{Convergence Property of LPGE}}
The following theorem reveals that the proposed lightweight policy gradient estimation is provably convergent and shows its convergence property. 
\begin{theorem}
\label{thm:convergence} 
% \red{you need to write a formal one and the corresponding assumptions. on $\Phi(\bs{s})$} We
Assuming that $\Phi(\mathbf{S})$ is L-smooth and the policy gradient variance $\mathbb{E}\lVert\mathcal{L}_{\mathcal{D}_{val}}\left({\theta}_1(\mathbf{M})\right) \nabla_{\mathbf{S}} \ln p(\mathbf{M}|\mathbf{S})-\nabla_{\mathbf{S}}\Phi(\mathbf{S})\rVert^2\leq\sigma^2$,we denote the gradient mapping $\mathcal{G}^t$ at $t$-th iteration as 
% \begin{equation*}
    $$\mathcal{G}^t=\frac{1}{\eta}\left(\mathbf{S}^t - \mathcal{P}_{\mathcal{C}}\left(\mathbf{S}^t-\eta \nabla_{\mathbf{S}}\Phi(\mathbf{S}^t)\right)\right).$$
% \end{equation*}
Based on the above assumptions and setting the step size $\eta<\frac{1}{L}$, we can have following conclusion:
$$\frac{1}{T} \sum_{t=1}^T \mathbb{E}\left\|\mathcal{G}^t\right\|^2 \leq \frac{8-2 L \eta}{2-L \eta} \sigma^2,$$ 
when $T\rightarrow \infty$.    
\end{theorem}
The proof for the theorem could be found in the appendix~\ref{sec:appendix-proof}.
% \subsection{Discussion}
% \label{subsec:discussion}
% \textbf{Coreset methods.} rely on heuristic criteria to select a subset from observed interactions, failing to guarantee the optimal solution for downstream recommendation task. Distinct from them, DConRec generate interactions
% \input{sections/sub-sections/append-proof}

% The proof is provided in the appendix~\ref{sec:appendix-proof}.
% The proof the above theorem could be found in~\cite{PCoreset-icml2022}.\red{you could introduce the proof if this is space left.}

% \red{Under the one-step optimization scheme, the mathematical of ${\theta}^*$ should be re-written as ${\theta}_0$ and you do it. you should explain how  Your proof for is different from the one in~\cite{doscond-kdd2022}}. Before stating your theorem, you need to list three assumptions: the loss function on original datasets is convex,  the loss function on original datasets is L-smooth and $\|\theta\|^2\leq M^2 (M>0)$.

% \input{sections/sub-sections/algorithms}
% \input{sections/sub-sections/figures}

%% file: sections/sub-sections/fig-datapoolAndOSPGE.tex
\begin{figure*}[t]
\vskip -0.1in
\centering
% \subfloat[Pre-augmentation]{\label{fig:datapool-gen}\includegraphics[width=0.31\linewidth]{figures/data-pool-gene7.pdf}}\hspace{0.04\linewidth}
% \subfloat[Condensing Recommendation Data]{\label{fig:os-pge}\includegraphics[width=0.58\linewidth]{figures/pdc-framework-20.pdf}}
\includegraphics[width=0.95\linewidth]{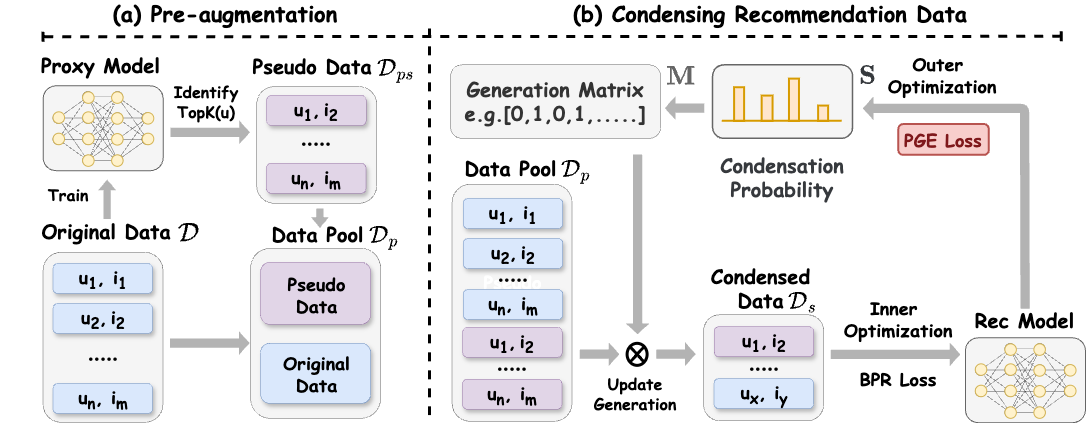}
\vskip -0.1in
\caption{(a) Generate data pool via pre-augmentation, where pseudo data is the set of items that have not interacted with user $u$ but they are ranked top-K for user $u$ by proxy model, identified as potential interests of user $u$. (b) Generate the condensed data $\mathcal{D}_s$ based on the pre-augmented recommendation dataset (\textit{data pool}: $\mathcal{D}_p$) via lightweight policy gradient estimation.}
\label{fig:pdc-framework-combine}
\vskip -0.15in
\end{figure*}

%% file: sections/experiments.tex
\section{Experiments}
% In this se
% In this section, we evaluate DConRec mainly from following aspects: (\textbf{RQ1}) How is the quality of dataset condensed by \ournameAbbr, as reflected in the performance of models trained on them? (\textbf{RQ2}) How efficient is \ournameAbbr? (\textbf{RQ3}) Can DConRec preserves the preference information of users in different groups? (\textbf{RQ4}) Can DConRec work well with different architectures of recommendation models? (\textbf{RQ5}) How does the choice of proxy models for pre-augmentation influence dataset quality? (\textbf{RQ6}) What is the performance of DConRec across various condensation ratios? \textbf{Besides}, we also conduct other studies to investigate the characteristics of DConRec. %\red{check}
In this section, we evaluate DConRec from these aspects: 
\begin{itemize}[leftmargin=*]
    \item {RQ1}: How is the quality of condensed dataset? 
    \item {RQ2}: How efficient is DConRec for condensation? 
    \item {RQ3}: Can DConRec preserve diverse user preferences? 
    \item {RQ4}: Can DConRec work well while adopting different recommendation models as backbone? 
    \item {RQ5}: Are the condensed datasets generalizable to train different recommendation models?
    \item {RQ6}: How does the choice of proxy models for pre-augmentation influence qualities of condensed datasets? 
    \item {RQ7}: How does condensation ratio influence the quality of condensed datasets? 
\end{itemize} %\red{check}
% {Besides}, we also further launch studies on DConRec.
\label{sec:experiments}
\subsection{Experimental Settings}
\textit{\textbf{Datasets.}} We evaluate the performance of our proposed framework on several real-world datasets: \textbf{Ciao}~\cite{ciao2012dataset}, \textbf{Gowalla}~\cite{gowalla2011kdd}, \textbf{Dianping}~\cite{dianping2015RecSys} and \textbf{ML-20M}~\cite{acm2015ml-20m}. For each dataset, we divide the interactions into 80\%/10\%/10\% for training/validation/testing. We conduct condensation/selection on the training set and train models on condensed datasets. Finally, we evaluate the performance on testing set. The statistics of the datasets can be found in Table~\ref{tab:dataset-statistics}.
\input{sections/sub-sections/experiments/tab-datasets}

\textit{\textbf{Baselines.}} We compare our method with baselines that could be implemented on discrete interaction data: three selection-based methods (Random, Majority and SVP-CF~\cite{iclr2020SVPOriginal,wsdm22SVP-data-gene}) and the condensation-based methods (GradMatch~\cite{doscond-kdd2022,zhaoICLR2021DC,ctrDD2023RecSys} and Distill-CF~\cite{infiniteRec22Nips}): 
\begin{itemize}[leftmargin=*]
    \item {Random} randomly samples a portion of interactions from the training set to form the small datasets for model training.
    \item {Majority} samples interactions from users with lower degrees, thereby maximizing the inclusion of users within the dataset.
    \item {SVP-CF}\cite{wsdm22SVP-data-gene} is a proxy-based coreset selection strategy for collaborative filtering datasets. We adopt MF as the proxy model.
    \item {GradMatch} is a prevalent condensation technique via matching the gradients of model parameters on condensed data and original data. We follow Doscond~\cite{doscond-kdd2022} and GCM~\cite{ctrDD2023RecSys} to implement the \textit{efficient} \textit{one-step gradient matching scheme} for the data synthesis.
    \item {{Distill-CF}}~\cite{infiniteRec22Nips} A NTK-based data condensation method for creating summaries by synthesizing fake users and their corresponding interactions. Since we aims at condensing interaction sets in our experiment, we set the product of the number of fake users and their interactions to the desired condensation size for fair comparison. We utilize its original implementation for evaluation.
\end{itemize}
%\red{the implementation detials of them could be found in appendix. since it it devised forxxx, we re-implement it on xxx}

\textit{\textbf{Backbone Models.}} In this paper, we conduct the evaluation on DConRec with following recommendation models:
\begin{itemize}[leftmargin=*]
    \item {MF}~\cite{rendle-et-al:bpr2009} is a traditional matrix factorization method that uses pairwise ranking loss based on implicit feedback.
    \item {NGCF}~\cite{sigir2019NGCF} is a graph-based CF method largely follows the standard GCN~\cite{kipf2016gcn}. It additionally encodes the second-order feature interaction during message passing.
    \item {LightGCN}~\cite{he-et-al:lightgcn} is a state-of-the-art recommendation model based on GNNs, which improves performance by removing activation and feature transformation. 
    \item {XSimGCL}~\cite{XSimGCL2022TKDE} is a state-of-the-art recommendation model enhanced by self-supervised learning (SSL), which boosts the performance by a cross-layer contrastive learning.
\end{itemize}

\textit{\textbf{Evaluation Protocol.}} We test the top-K ranking performance of models trained on the condensed datasts, reporting the \textit{Recall} and \textit{NDCG}, which we denote by \textit{R@K} and \textit{N@K}. It involves three stages: (1) condensing/selecting a small dataset, (2) training a model on the small dataset and (3) test the model's performance.

\input{sections/sub-sections/experiments/tab-overall-compare}

\textit{\textbf{Implementation Details.}} For the sake of universality, we adopt MF as the backbone model for overall performance comparison in Table~\ref{tab:overall-com}. The proxy model defaults to MF unless stated otherwise. When assessing cross-architecture performance and condensed dataset generalizability, we substitute MF with NGCF, LightGCN, and XSimGCL, setting the layer number to 2 or 3. For XSimGCL, the SSL weight is independently tuned for each dataset. The embedding dimension is set to 64. 
% Unless otherwise stated, the evaluation is conducted while condensation ratio $r$ is 0.25. 
The learning rate for condensation ranges $\{0.1,0.01,0.05,0.005,0.0005\}$ and the decay is 10 per 100 epochs, bounded by 1e-4. The pseudo data ratio $r_{ps}$ ranges from 0.1 to 0.9. 
% The outer training epochs is set to 1000. 
Model training utilizes the Adam optimizer with various initial learning rates, applying early stop. The environment includes a PyTorch (Version 1.12) on an Nvidia A30 GPU (Memory: 24GB, Cuda version: 11.3).% Parameter settings.
%The implementation details is provided in appendix~\ref{sec:appendix-exp-details}.
\input{sections/sub-sections/experiments/tab-efficiency-comare}
\input{sections/sub-sections/experiments/tab-head-tail-compare}
\subsection{Performance Comparison}
{\textbf{{Quality of Condensed Datasets (RQ1).}}}
To evaluate the quality of the datasets condensed by DConRec, we measure the performance of MF trained on them. Specifically, with the condensation ratio $r$ is set to $0.25$, We report the metric \textit{Recall@K} and \textit{NDCG@K}, where K is set to 5 and 10. For the baselines, the condensation ratio is also set to $0.25$. The results are summarized in Table~\ref{tab:overall-com}, from which we have following observations:
\begin{itemize}[leftmargin=*]
    \item {Superiority of DConRec over baselines:} Our proposed \textit{\ournameAbbr} outperforms the baselines across different datasets. Even when the size of datasets are reduced by 75\%, we could achieve over comparable performance of the original datasets. Notably, on Dianping, we approximate 98\% of the original performance with only 25\% data. By contrast, baselines underperform our method by a large margin. These results demonstrate the effectiveness of our method in preserving the information of the original datasets.
    % \item  Increasing the size of condensed datasets can improve the performance for all the methods on different datasets. This is quite easy to follow since more data samples indicate more information preserved in small datasets. Interestingly, we found that when the condensation ratio is raised to 0.5, the performance of our proposed method on Dianping even outperform the performance with original datasets. This phenomenon can be attributed to the fact that our method ameliorates the deleterious impact of false negative items, thereby effectively denoising the original dataset.
    \item {Direct application of GradMatch to recommendation may lead to unsatisfactory results:} We observe that the GradMatch does not perform well. We argue that this is due to its design tailored for data with continuous features (e.g., images, node features, etc). Thus, direct application of it in generating interactions cannot capture adequate information about users' preference for condensed dataset. Therefore, it cannot even beat Majority and SVP-CF on three datasets, both of which are heuristic selection methods and they can capture partial preference information.
\end{itemize}

\noindent{\textbf{Condensing Efficiency ({RQ2}).} }
Since we also aim to achieve an efficient condensation method, we evaluate the running-time of DConRec and compare it with GradMatch. Further, to examine the efficiency of the proposed optimization, we also compare a variant of \ournameAbbr, which only counts the time of optimizing dataset without the time for inner model training. We denote the variant by \ournameAbbr-Optim. Under the same setting of Table~\ref{tab:overall-com}, We run 100 epochs for boths methods and record the time. The result is summarized in Table~\ref{tab:efficiency-compare}. We can observe that, in terms of updating dataset, DConRec can be $8\times$ faster than GradMatch, which verifies the efficiency of our proposed method. Further, via removing the time of inner training, we can find that the real optimization time of DConRec is further shorter. The efficiency of DConRec stems from its design of forward calculation (PGE) to update the synthetic dataset, avoiding the time-consuming process of backpropagation, which is required in the gradient matching scheme of GradMatch.

\noindent{\textbf{Performance on Different User Groups (RQ3).}} We investigate the ability of DConRec in preserving the information for different groups of users and we report the results on four datasets in Table~\ref{tab:head-tail-performance}. Specifically, we split the users into three groups: Head (100 \textless num. of interections), Torso (10 \textless num. of interections \textless 100) and Tail (num. of interections \textless 10). Here, we adopt MF as the backbone model and the test model. From Table~\ref{tab:head-tail-performance}, we can observe that our proposed method can preserve the preference information in most cases for users in all three groups
% and thus yields the best performance 
when compared to other baselines, close to the performance of model trained on full data. In particular, while other methods %fail to achieve this 
suffer significant performance decrease on the tail user group, our method still perform excellently comparable.

\noindent{\textbf{Training Efficiency.}}
Due to the large size of the real-world recommendation datasets, it is quite expensive to train the models based on them. In this part, we aim to examine the training efficiency of the recommender models on our condensed dataset. We record the training curves of MF and LightGCN on condensed Dianping dataset and original Dianping dataset. As shown in Figure~\ref{fig:training-efficiency-MF} and~\ref{fig:training-efficiency-LG}, we can observe that training on the condensed dataset is greatly more efficient than training on the original dataset. Notably, the training on the condensed dataset is more efficiency (speedup over $6\times$) with comparable performance.
% \input{sections/sub-sections/experiments/fig-training-efficiency} 
% \input{sections/sub-sections/experiments/fig-trainEfficiency-condenRatios}
\input{sections/sub-sections/experiments/tab-cross-architects}
\subsection{Generalization of DConRec}
In this subsetion, we examine the generalizability of the condensed datasets to train different models and the generalizability of DConRec cross different architectures of backbone model on the four datasets. 

%We report the results on dataset Ciao in Table~\ref{tab:cross-architecture}. \\
{\textbf{Different Architectures of Model as Backbone (RQ4).}} The proposed DConRec is highly flexible since we can adopt various architectures for the backbone model. We investigate the performances of DConRec when equipping it with different backbone models (i.e., MF, NGCF, LightGCN and XSimGCL). From Table~\ref{tab:cross-architecture}, we can see that the dataset condensed by different mdoels all show strong generalizability on other architectures. Interestingly, we can observe that XSimGCL yields the best performance regarding the dataset condensed by the DConRec with the same backbone model while setting XSimGCL to be the backbone model may not produce the best testing performance regarding the same test model.

{\textbf{Different Architectures of Model for Testing (RQ5).}} Specifically, we show the test performance when using dataset condensed by one model to train different recommenders. As depicted in Table~\ref{tab:cross-architecture}, we choose MF~\cite{rendle-et-al:bpr2009}, NGCF~\cite{sigir2019NGCF}, LightGCN~\cite{he-et-al:lightgcn} and XSimGCL~\cite{XSimGCL2022TKDE} to evaluate the quality of the condensed datas, observing that the dataset condensed by one specific model is also informative to train models of other architectures. Among them, we can observe that XSimGCL outperforms other models regarding the same condensed datasets under most settings, consistent with the performances trained on the original datasets.
% \subsection{Effects on Different User Groups (RQ4)}
% We investigate the ability of our proposed DConRec in preserving the information for different groups of users and we report the results on three datasets in Table~\ref{tab:head-tail-performance}. Specifically, we split the users into three groups: Head (num. of interections \textgreater 100), Torso (100 \textgreater num. of interections \textgreater 10) and Tail (10 \textgreater num. of interections). Here we adopt MF as the backbone model and the test model. As shown in Table~\ref{tab:head-tail-performance}, we can observe that our proposed method can preserve the information for users in all three groups when compared to other baselines, close to the performance of model trained on full data. Particularly, while other methods fail to preserve the information for long-tailed users during condensation, our method still perform excellently comparable. 
\input{sections/sub-sections/experiments/fig-different-condensed-ratios}

\input{sections/sub-sections/experiments/tab-proxy-models}

\subsection{Proxy Models Study in Pre-augmentation (RQ6)}
In this section, we investigate the effects of adopting different proxy models for DConRec to produce the condensed datasets. In Table~\ref{tab:proxy-model-study}, we report the performances of varying the proxy models among MF~\cite{rendle-et-al:bpr2009}, NGCF~\cite{sigir2019NGCF}, LightGCN~\cite{he-et-al:lightgcn} and XSimGCL~\cite{XSimGCL2022TKDE}, reporting the testing evaluation with model MF. We can observe that DConRec can work well with different architectures of proxy model. Interestingly, more powerful proxy model can contribute to better performance (e.g., adopting XSimGCL as the proxy model yields the best performance on these four datasets). We argue that a more powerful proxy model can better capture preferences, thereby more effectively including potential user interests.

\subsection{Varying Condensation Ratios (RQ7)}
\label{subsec:rq7-conden-ratio}
In this section, we evaluate DConRec under different condensation ratios, ranging from $0.1$ to $0.9$. We report the results regarding Recall@5 and NDCG@5 in Figure~\ref{fig:diff-condensed-ratios}. 

\textbf{Results.} We can observe that the performance is roughly proportional to the size of the condensed dataset since larger size denotes more information from original dataset. 
As depicted from Figure~\ref{fig:diff-condensed-ratios}, at most cases (even when the condensation ratio is $0.1$), the datasets condensed by our method are qualified and the performance of model trained on them is close to those trained on full datasets. 

\textbf{Discussion.} We attribute the decent performance to the ability of DConRec in preserving the potential preference of users since the key to successful recommendation is modeling personalized preference. Furthermore, since not all interactions contribute positively to preference modeling, their reduction during condensation may not hinder personalized interests modeling.

\input{sections/sub-sections/experiments/tab-ablation-study-module}

\subsection{Further Investigations}
In this section, we further explore following questions: (1) What is the effect of pre-augmentation? (2) How does performance vary with different sizes of pseudo data? (3) How do varying inner training epochs affect performance? (4) Can embeddings learned from condensed datasets replicate the patterns of those from the full dataset?

{\textbf{Ablation Study on Pre-augmentation.}}
To investigate the effect of pre-augmentation on the performance of DConRec, we study the variant of our method by removing the module of pre-augmentation, indicated by \textbf{w\textbackslash o-PA} in Table~\ref{tab:ablation-study}. The results regarding Recall@10 and NDCG@10 on four datasets are summarized in Table~\ref{tab:ablation-study} and we can see that the elimination of pre-augmentation lead to great performance degradation. Besides, in Figure~\ref{fig:intro-data-qualities}, we can also observe that while directly training on the pre-augmented original dataset, the performance improvement is limited, which indicate that the pre-augmentation specifically benefit the condensation quality. This also demonstrates the necessities of pre-augmentation in preserving the user preference.% from the original interaction set for synthetic dataset. 
\input{sections/sub-sections/experiments/fig-multi-parameters}

{\textbf{Varying Sizes of Pseudo Dataset in Pre-augmentation.}}
We study how different sizes of pseudo dataset affect the performance. Concretely, we vary the pseudo data ratio $r_{ps}$ from 0.1 to 0.9. As the reported results on dataset Ciao in Figure~\ref{fig:ciao-pseudo-ratio-recall}, we find that more pseudo data does not invariably yield better performance, some times even worse. We posit that this arises from the fact that a significant portion of the pseudo data might introduce noise and a larger size of parameters to be optimized for the generation of condensed dataset, in line with the discussion in section~\ref{subsec:rq7-conden-ratio}.

{\textbf{Investigation on the Lightweight Update Scheme.}}
To further investigate the effect of lightweight update scheme, we vary the inner epochs from 1 to 10 and test the performance on dataset Ciao. As illustrated in Figure~\ref{fig:dianping-pseudo-ratio-recall}, we can observe that as the number of inner model training increase, the performance of DConRec does not vary a lot. This verifies the effectiveness of the lightweight design and it is in line with the results in previous studies~\cite{doscond-kdd2022,KDD23bilevelDenoise}.

{\textbf{Visualization}}
To further examine the quality of the condensed datasets synthesized by our method, we utilize t-SNE~\cite{JMLR08tSNE} to visualize the learned user and item embeddings. We train MF on datasets produced by \textit{Random}, \textit{DConRec} and \textit{Full Dataset}. We set the condensation ratio to 0.25 and provide the t-SNE plots of the learned embeddings for Ciao in Figure~\ref{fig:t-sne-visualization}. It is observed that embeddings learned by randomly selecting interactions deteriorate the patterns in original dataset, which changed the personalization of users. In contrast, the embeddings learned by our method preserve similar patterns to the one with full dataset. This makes the comparable performance of DConRec explainable. 
\input{sections/sub-sections/experiments/fig-visualization}

%% file: sections/sub-sections/experiments/tab-datasets.tex
\begin{table}[]
\vskip -0.15in
\caption{Dataset statistics.}
\vskip -0.15in
\centering
\scalebox{1.}{\begin{tabular}{r|rrrr}
\toprule%[1.0pt]
Dataset  & \#User & \#Item  & \#Interactions & Density \\ \midrule
Ciao     & 7,375  & 105,114 & 284,086        & 0.00037 \\
Gowalla  & 29,859 & 40,989  & 1,027,464      & 0.00084 \\
Dianping & 16,396 & 14,546  & 51,946         & 0.00022 \\ 
ML-20M & 138,493 & 27,278 & 20,000,263&0.00529\\
\bottomrule%[1.0pt]
\end{tabular}}
\label{tab:dataset-statistics}
\vskip -0.15in
\end{table}

%% file: sections/sub-sections/experiments/tab-overall-compare.tex
\begin{table*}[htbp]
% \vskip -0.15in
\centering
\caption{Overall performance comparison.``\textit{Full Dataset}" indicates the performance of models trained on the original datasets.``\textit{R@K}" denotes Recall@K and``\textit{N@K}" denotes NDCG@K.``\textit{Sel-based}" denotes selection based methods and ``\textit{Cond-based}" denotes dataset condensation methods.}% $\Delta$
\vskip -0.1in
\scalebox{1.}{
\begin{tabular}{cc|cc|cc|cc|cc}
\toprule%[1.2pt]
\multicolumn{2}{c|}{Dataset}      & \multicolumn{2}{c|}{Ciao}  & \multicolumn{2}{c|}{Gowalla}   & \multicolumn{2}{c|}{Dianping}  & \multicolumn{2}{c}{ML-20M}  \\ \midrule
\multicolumn{2}{c|}{Metrics} & R@10       & N@10 & R@10  & N@10         & R@10 & N@10 & R@10 & N@10         \\ \midrule
\multicolumn{1}{c|}{\multirow{3}{*}{\begin{tabular}[c]{@{}c@{}}Sel\\ -based\end{tabular}}}    & Random    & 0.0092 & 0.0076 & 0.0457 & 0.0327 &  0.0169  & 0.0091  &0.0838 &0.1050        \\
\multicolumn{1}{c|}{}       & Majority  & 0.0180 & 0.0117 &  0.0667 & 0.0404 & 0.0236 &  0.0120  &0.1176 &0.1103  \\
\multicolumn{1}{c|}{}       & SVP-CF       & 0.0289  & 0.0187 &  0.0836  & 0.0567 & 0.0523 & 0.0285   &0.1572 &0.1825    \\ \midrule
\multicolumn{1}{c|}{\multirow{3}{*}{\begin{tabular}[c]{@{}c@{}}Cond\\ -based\end{tabular}}} & GradMatch & 0.0124 & 0.0092 &  0.0548 & 0.0394 & 0.0200 & 0.0111  &0.1519 &0.1789   \\
\multicolumn{1}{c|}{}       & {Distill-CF}   & 0.0169 & 0.0199 &  -- & -- &  0.0191 & 0.0123 & -- & --   \\ 
\multicolumn{1}{c|}{}       & \textbf{DConRec}   & \textbf{0.0302} & \textbf{0.0200} &  \textbf{0.0930} & \textbf{0.0651} &  \textbf{0.0614} & \textbf{0.0340} &\textbf{0.1702} &\textbf{0.2021}   \\ \midrule
% \multicolumn{2}{c|}{\cellcolor[HTML]{EFEFEF}Quality} & \cellcolor[HTML]{EFEFEF}91.24\%         & \cellcolor[HTML]{EFEFEF}96.71\%         & \cellcolor[HTML]{EFEFEF}91.57\%         & \cellcolor[HTML]{EFEFEF}94.34\%         & \cellcolor[HTML]{EFEFEF}91.61\%         & \cellcolor[HTML]{EFEFEF}91.63\%         & \cellcolor[HTML]{EFEFEF}90.11\%         & \cellcolor[HTML]{EFEFEF}90.42\%         & \cellcolor[HTML]{EFEFEF}99.26\%         & \cellcolor[HTML]{EFEFEF}99.80\%         & \cellcolor[HTML]{EFEFEF}97.56\%         & \cellcolor[HTML]{EFEFEF}98.19\%         \\ \midrule
\multicolumn{2}{c|}{Full Dataset}       & 0.0312 & 0.0212 & 0.1015  & 0.0720 & 0.0615 & 0.0346  &0.2031 &0.2414  \\ \bottomrule%[1.2pt]
\end{tabular}
}
\label{tab:overall-com}
\vskip -0.15in
\end{table*}

%% file: sections/sub-sections/experiments/tab-efficiency-comare.tex
% \begin{table}[]
% \caption{Comparison of the running time for 100 epochs. \textit{DConRec-Optim} denotes the time of optimizing the synthesized dataset, without the one-step update for inner model training.}
% \centering
% \scalebox{0.75}{
% \begin{tabular}{c|ccc|ccc}
% \toprule
% Cond. Ratio   & \multicolumn{3}{c|}{0.25}     & \multicolumn{3}{c}{0.5}       \\ \midrule
% Dataset       & Ciao    & Gowalla  & Dianping & Ciao    & Gowalla  & Dianping \\ \midrule
% DosCond       & 8min43s & 16min44s & 35min56s & 17min2s & 33min53s & 51min49s \\
% DConRec       & 3min22s & 13min11s & 30min19s & 3min48s & 16min04s & 32min22s \\
% DConRec-Optim & 0min52s & 4min11s  & 10min19s & 1min18s & 4min24s  & 10min42s \\ \bottomrule
% \end{tabular}
% }
% \label{tab:efficiency-compare}
% \end{table}
\begin{table}[]
\vskip -0.15in
\centering
% \vskip -0.12in
\caption{Comparison of the running time for 100 epochs.``{DConRec-{Op}}" denotes the time of optimizing the synthesized dataset via policy gradient matching, without the time for inner model update.}
\vskip -0.1in
\scalebox{1.}{
\begin{tabular}{c|cccc}
\toprule%[1.0pt]
% Methods       & Ciao    & Gowalla  & Dianping & ML-20M\\ \midrule
% GradMatch       & 8min43s & 16min44s & 35min56s & 162min47s\\
%  DConRec       & 3min22s & 13min11s & 30min19s & 143min38s \\
% DConRec-Op & 0min52s & 4min11s  & 10min19s &17min06s\\ 
                      % \midrule
Methods & GradMatch & DConRec & DConRec-Op \\ \midrule
Ciao & 8min43s&3min22s &\textbf{0min52s} \\
Gowalla &16min44s &13min11s & \textbf{4min11s}\\
Dianping &35min56s &30min19s &\textbf{10min19s} \\
ML-20M &162min47s &143min38s &\textbf{17min06s} \\

% \multirow{3}{*}{0.5}  & DosCond       & 17min2s & 33min53s & 51min49s \\
%                       & DConRec       & 3min48s & 16min04s & 32min22s \\
%                       & DConRec-Optim & 1min18s & 4min24s  & 10min42s \\
\bottomrule%[1.0pt]
\end{tabular}
}
\label{tab:efficiency-compare}
\vskip -0.15in
\end{table}

%% file: sections/sub-sections/experiments/tab-head-tail-compare.tex
% Please add the following required packages to your document preamble:
% \usepackage{multirow}
% \usepackage[table,xcdraw]{xcolor}
% Beamer presentation requires \usepackage{colortbl} instead of \usepackage[table,xcdraw]{xcolor}
\begin{table*}[]
\centering
\vskip -0.15in
\caption{Performance comparison on different user groups. The bold denotes the best performance with the condensed data.}
\vskip -0.12in
\scalebox{0.95}{\begin{tabular}{c|c|cc|cc|cc|cc}
% \toprule%[1.2pt]
\toprule%[1.2pt]
\multirow{2}{*}{Users} &\multirow{2}{*}{Methods}&\multicolumn{2}{c|}{Ciao}&\multicolumn{2}{c|}{Gowalla}&\multicolumn{2}{c|}{Dianping}&\multicolumn{2}{c}{ML-20M}  \\ %\cmidrule{3-10} 
 && R@10&N@10&R@10&N@10&R@10&N@10&R@10&N@10         \\ \midrule

\multirow{4}{*}{Head}  &SVP &\textbf{0.0202}&\textbf{0.0372}&0.0468&0.0914&0.0306&0.0463&0.1144&0.2378\\
&GradMatch& 0.0090 &0.0222&0.0420&0.0794&0.0227&0.0349&0.1159&0.2283\\
&\textbf{DConRec}&{0.0183}&{0.0309}&\textbf{0.0343}&\textbf{0.0547}&\textbf{0.0397}&\textbf{0.0605}&\textbf{0.1205}&\textbf{0.3022}\\
&Full Data& 0.0213&0.0438&0.0579&0.1168&0.0389&0.0617&0.1341&0.3256\\ \midrule

\multirow{4}{*}{Torso} &SVP    &0.0269&\textbf{0.0144}&0.1002&0.0534&0.053 &0.0269&0.0317&0.0183\\
&GradMatch& 0.0101&0.0063&0.0608&0.0373&0.0169&0.0082&0.0288&0.0139\\
&\textbf{DConRec}&\textbf{0.0257}&{0.0138}&\textbf{0.115}&\textbf{0.065}&\textbf{0.0632}&\textbf{0.0320}&\textbf{0.0390}&\textbf{0.0193}\\
&Full Data& 0.0296&0.0164&0.1211&0.0680 &0.0609&0.0315&0.0741&0.0399\\ \midrule

\multirow{4}{*}{Tail}  &SVP    &0.0304&0.0190 &0.0849&0.0558&0.0501&0.0304&0.1812&0.1414\\
&GradMatch& 0.0116&0.0089&0.0608&0.0406&0.0252&0.01500&0.1856&0.1474\\
&\textbf{DConRec}&\textbf{0.0306}&\textbf{0.0205}&\textbf{0.0946}&\textbf{0.0634}&\textbf{0.0587}&\textbf{0.0355}&\textbf{0.2028}&\textbf{0.1560}\\
&Full Data& 0.0329&0.0220 &0.1024&0.0697&0.0634&0.0392&0.2328&0.1844\\ 
                         \midrule

\multirow{4}{*}{Overall}&SVP    &0.0289&0.0187&0.0836&0.0567&0.0523&0.0285&0.1572 &0.1825\\
&GradMatch& 0.0124&0.0092&0.0548&0.0394&0.0200  &0.0111&0.1519 &0.1789\\
&\textbf{DConRec}&\textbf{0.0302}&\textbf{0.0200} &\textbf{0.0930}&\textbf{0.0651}&\textbf{0.0614}&\textbf{0.0340}&\textbf{0.1702} &\textbf{0.2021}\\
&Full Data& 0.0312 &0.0212&0.1015&0.0720 &0.0615&0.0346&0.2031 &0.2414\\ 
                         \bottomrule%[1.2pt]
\end{tabular}}
\label{tab:head-tail-performance}
\vskip -0.17in
\end{table*}

%% file: sections/sub-sections/experiments/tab-cross-architects.tex
% Please add the following required packages to your document preamble:
% \usepackage{multirow}
\begin{table}[]
% \vskip -0.05in
\vskip -0.1in
\centering
\caption{Generalization of DConRec: equip it with different backbone models (RQ4) and utilize the condensed data to train various models (RQ5).}
\vskip -0.15in
\scalebox{0.95}{
\begin{tabular}{c|c|cc|cc}
\toprule
\multirow{2}{*}{Backbone} & \multicolumn{1}{c|}{\multirow{2}{*}{Test Model}} & \multicolumn{2}{c|}{Ciao}     & \multicolumn{2}{c}{Gowalla} \\
                          & \multicolumn{1}{c|}{} & R@10   & N@10   & R@10   & N@10   \\ \midrule
\multirow{4}{*}{MF}       & MF                    & 0.0302 & 0.0200   & 0.0930  & 0.0651 \\
                          & NGCF                  & 0.0282 & 0.0179 & 0.0891 & 0.0624 \\
                          & LightGCN              & 0.0300   & 0.0201 & 0.0914 & 0.0644 \\
                          & XSimGCL               & 0.0293 & 0.0194 & 0.0812 & 0.0565 \\ \midrule
\multirow{4}{*}{LightGCN} & MF                    & 0.0289 & 0.0186 & 0.0932 & 0.0652 \\
                          & NGCF                  & 0.0282 & 0.0185 & 0.0893 & 0.0623 \\
                          & LightGCN              & 0.0299 & 0.0197 & 0.0915 & 0.0643 \\
                          & XSimGCL               & 0.0298 & 0.0195 & 0.0898 & 0.0624 \\ \midrule
\multirow{4}{*}{NGCF}     & MF                    & 0.0289 & 0.0184 & 0.0920  & 0.0643 \\
                          & NGCF                  & 0.0279 & 0.0186 & 0.0868 & 0.0611 \\
                          & LightGCN              & 0.0295 & 0.0195 & 0.0919 & 0.0644 \\
                          & XSimGCL               & 0.0299 & 0.0193 & 0.0890  & 0.0621 \\ \midrule
\multirow{4}{*}{XSimGCL}  & MF                    & 0.0299 & 0.0190  & 0.0914 & 0.0639 \\
                          & NGCF                  & 0.0277 & 0.0185 & 0.0884 & 0.0618 \\
                          & LightGCN              & 0.0299 & 0.0197 & 0.0912 & 0.0639 \\
                          & XSimGCL               & 0.0295 & 0.0930  & 0.0805 & 0.0563 \\ \bottomrule\toprule
\multirow{2}{*}{Backbone} & \multicolumn{1}{c|}{\multirow{2}{*}{Test Model}} & \multicolumn{2}{c|}{Dianping} & \multicolumn{2}{c}{ML-20M}  \\
                          & \multicolumn{1}{c|}{} & R@10   & N@10   & R@10   & N@10   \\ \midrule
\multirow{4}{*}{MF}       & MF                    & 0.0606 & 0.0335 & 0.1702 & 0.2021 \\
                          & NGCF                  & 0.0602 & 0.0332 & 0.1648 & 0.1986 \\
                          & LightGCN              & 0.0377 & 0.0202 & 0.1529 & 0.1824 \\
                          & XSimGCL               & 0.0545 & 0.0298 & 0.1716 & 0.2055 \\ \midrule
\multirow{4}{*}{LightGCN} & MF                    & 0.0605 & 0.0334 & 0.1698 & 0.2017 \\
                          & NGCF                  & 0.0606 & 0.0334 & 0.1645 & 0.1980  \\
                          & LightGCN              & 0.0378 & 0.0201 & 0.1527 & 0.1822 \\
                          & XSimGCL               & 0.0540  & 0.0295 & 0.1749 & 0.2095 \\ \midrule
\multirow{4}{*}{NGCF}     & MF                    & 0.0609 & 0.0338 & 0.1501 & 0.1783 \\
                          & NGCF                  & 0.0607 & 0.0335 & 0.1655 & 0.1988 \\
                          & LightGCN              & 0.0296 & 0.0155 & 0.1749 & 0.2078 \\
                          & XSimGCL               & 0.0540  & 0.0297 & 0.1740  & 0.2086 \\ \midrule
\multirow{4}{*}{XSimGCL}  & MF                    & 0.0604 & 0.0334 & 0.1700   & 0.2017 \\
                          & NGCF                  & 0.0601 & 0.0333 & 0.1650  & 0.1983 \\
                          & LightGCN              & 0.0377 & 0.0204 & 0.1292 & 0.1558 \\
                          & XSimGCL               & 0.0537 & 0.0293 & 0.1749 & 0.2094 \\ \bottomrule
\end{tabular}
}
\label{tab:cross-architecture}
\vskip -0.25in
\end{table}

%% file: sections/sub-sections/experiments/fig-different-condensed-ratios.tex
\begin{figure*}[]
\vskip -0.1in
\centering
\subfloat[Ciao]{\label{fig:ciao-condensed-ratios}\includegraphics[width=0.465\columnwidth]{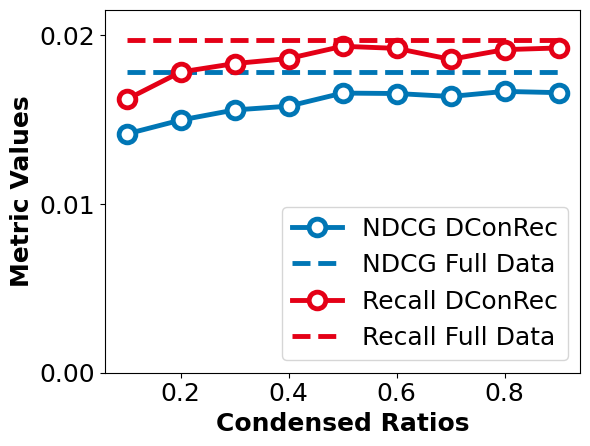}}
\hspace{0.01\columnwidth}
\subfloat[Gowalla]{\label{gowalla-condensed-ratios}\includegraphics[width=0.465\columnwidth]{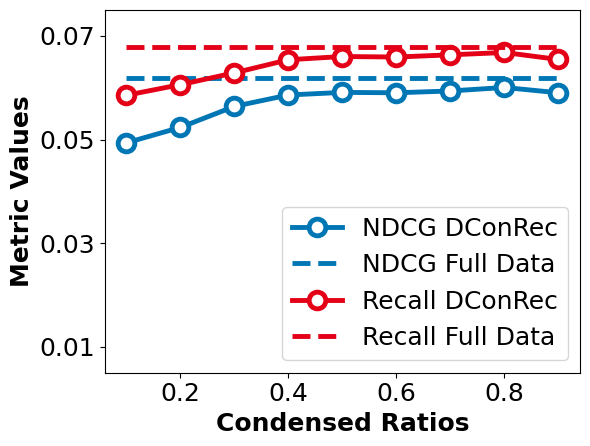}}
\hspace{0.01\columnwidth}
\subfloat[Dianping]{\label{fig:dianping-condensed-ratios}\includegraphics[width=0.45\columnwidth]{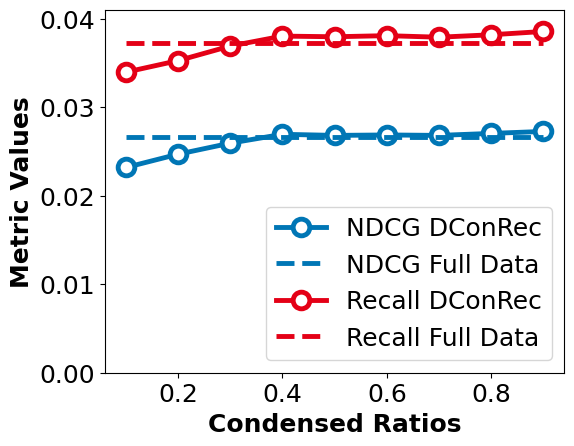}}
\hspace{0.01\columnwidth}
\subfloat[ML-20M]{\label{ml-20m-condensed-ratios}\includegraphics[width=0.465\columnwidth]{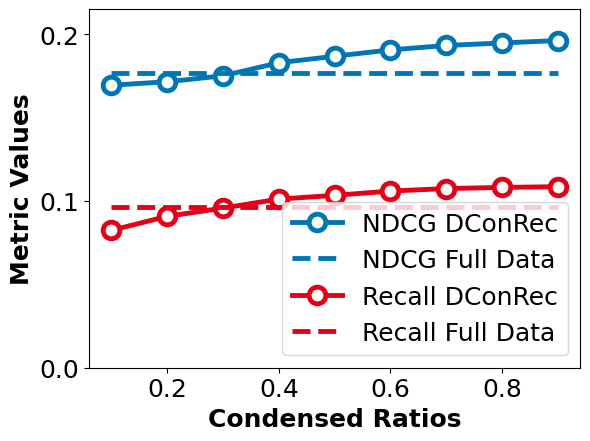}}
\vskip -0.1in
\caption{Vary condensed ratios. The solid lines denote the performance with condensed datasets, while the dashed lines indicate those with full ones.}
\label{fig:diff-condensed-ratios}
\vskip -0.2in
\end{figure*}

%% file: sections/sub-sections/experiments/tab-proxy-models.tex
\begin{table}[]
\vskip -0.05in
\caption{Effects of different architectures of poxy model.}
\vskip -0.1in
\scalebox{0.95}{
\begin{tabular}{c|c|cccc}
\toprule%[1.2pt]
Dataset                    & Proxy Model & R@5                       & R@10                     & N@5                         & N@10                        \\ \midrule
                           & MF          & 0.0180                          & 0.0302                        & 0.0163                         & 0.0200                         \\
                           & NGCF        & 0.0175                       & 0.0293                        & 0.0148                       & 0.01863                       \\
                           & LightGCN    & 0.0193                       & 0.0297                      & 0.0166                       & 0.01968                       \\
\multirow{-4}{*}{Ciao}     & XSimGCL     & 0.0193                       & 0.0328                        & 0.0167                       & 0.02083                       \\ \midrule
                           & MF          & 0.0622 & 0.0930 & 0.0558 & 0.0651 \\
                           & NGCF        & 0.0609                        & 0.0909                      & 0.0545                         & 0.06370                         \\
                           & LightGCN    & 0.0664                        & 0.0963                       & 0.0590                       & 0.0680                          \\
\multirow{-4}{*}{Gowalla}  & XSimGCL     & 0.0683                        & 0.0996                      & 0.0607                       & 0.07018                       \\ \midrule
                           & MF          & 0.0370                          & 0.0614                        & 0.0259                         & 0.03400                          \\
                           & NGCF        & 0.0369                        & 0.0606                       & 0.0256                         & 0.03340                         \\
                           & LightGCN    & 0.0396                       & 0.0639                        & 0.0279                         & 0.03595                        \\
\multirow{-4}{*}{Dianping} & XSimGCL     & 0.0407                       & 0.0657                        & 0.0287                        & 0.03688                       \\\midrule
                           & MF          & 0.1090                       & 0.1702                        & 0.2010                         & 0.2021                          \\
                           & NGCF        & 0.1086                        & 0.1703                       & 0.1998                        & 0.2012                         \\
                           & LightGCN    & 0.1099                       & 0.1721                        & 0.2027                         & 0.2034                        \\
\multirow{-4}{*}{{ML-20M}} & XSimGCL & 0.1105                       & 0.1750                        & 0.2088                        & 0.2107                       \\ \bottomrule%[1.2pt]
\end{tabular}
}
\label{tab:proxy-model-study}
\vskip -0.15in

\end{table}

%% file: sections/sub-sections/experiments/tab-ablation-study-module.tex
% Please add the following required packages to your document preamble:
% \usepackage{multirow}
\begin{table}[]
\centering
\vskip -0.05in

\caption{ Ablation study on the pre-augmentation.}
\vskip -0.1in

\scalebox{0.95}{
\begin{tabular}{c|c|cc}
\toprule%[1.2pt]
Dataset                   &Metrics   &  w\textbackslash{}o-PA &  DConRec         \\ \midrule
\multirow{2}{*}{Ciao}     %& R@5 &  0.0110                &  0.0120 \\
                          & R@10 &  0.0161                &  {0.0302} \\
                          %& N@5   &  0.0101                &  {0.0163}\\
                          & N@10   &  0.0115                &  {0.0200} \\ %\cmidrule{2-7} 
                          % & \multirow{2}{*}{0.5}  & R & 0.0227 & 0.0265                & 0.0288                 & {0.0309} \\
                          % &                       & N   & 0.0158 & 0.0182                & 0.0190                 & {0.0198} \\ 
                          \midrule
\multirow{2}{*}{Gowalla}  %& R@5 &  0.0517                &  {0.0622} \\
                          & R@10 &  0.0771                &  {0.0930} \\
                          %& N@5   &  0.0467                & {0.0558} \\
                          & N@10   &  0.0545                & {0.0651} \\ %\cmidrule{2-7} 
                          % & \multirow{2}{*}{0.5}  & R & 0.0935 & 0.0993                & 0.0917                 & {0.0996} \\
                          % &                       & N   & 0.0667 & 0.0706                & 0.0643                 & {0.0703} \\ 
                          \midrule
\multirow{2}{*}{Dianping} %& R@5 &  0.0164                &  {0.0370} \\
                          & R@10 &  0.0286                &  {0.0614} \\
                          %& N@5   &  0.0114                &  {0.0260} \\
                          & N@10   &  0.0155                &  {0.0340} \\ \midrule%\cmidrule{2-7} 
                          % & \multirow{2}{*}{0.5}  & R & 0.0348 & 0.0406                & 0.0602                 & {0.0638} \\
                          % &                       & N   & 0.0193 & 0.0224                & 0.0335                 & {0.0356} \\ 
\multirow{2}{*}{ML-20M} %& R@5 &  0.0164                &  {0.0370} \\
                          & R@10 &  0.1304                &  {0.1702} \\
                          & N@10   &  0.1569                &  {0.2021} \\ 
                          \bottomrule%[1.2pt]
\end{tabular}
}

\label{tab:ablation-study}
\vskip -0.2in

\end{table}

%% file: sections/sub-sections/experiments/fig-multi-parameters.tex
\begin{figure*}[]
% \vskip -0.1in
\centering
% \subfloat{\label{fig:ciao-threshold-study-ndcg}\includegraphics[width=0.495\columnwidth]{figures/experiments/ciao-threshold-tail.png}}\hspace{0.025\columnwidth}

% \subfloat[Training efficiency]{\label{fig:training-efficiency}\includegraphics[width=0.495\columnwidth]{figures/experiments/training-efficiency/MF-MF-MF-trainingEfficiency.png}}
\subfloat[Train curve-MF]{\label{fig:training-efficiency-MF}\includegraphics[width=0.495\columnwidth]{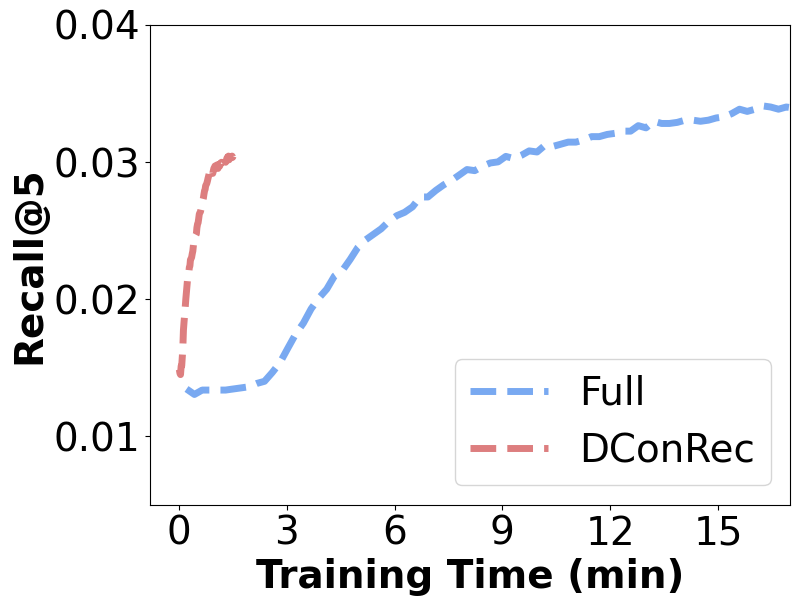}}
\subfloat[Train curve-LightGCN]{\label{fig:training-efficiency-LG}\includegraphics[width=0.495\columnwidth]{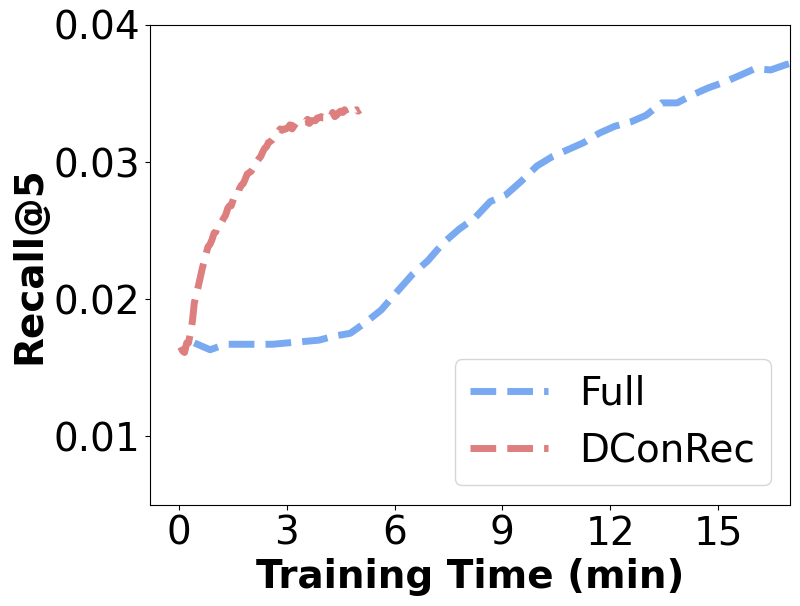}}
% \hspace{0.01\columnwidth}
% \subfloat[Varying condensed ratios]{\label{fig:diff-condensed-ratios}\includegraphics[width=0.495\columnwidth]{figures/experiments/different-condensed-ratios.png}}
\hspace{0.01\columnwidth}
\subfloat[Vary pseudo data ratio]{\label{fig:ciao-pseudo-ratio-recall}\includegraphics[width=0.495\columnwidth]{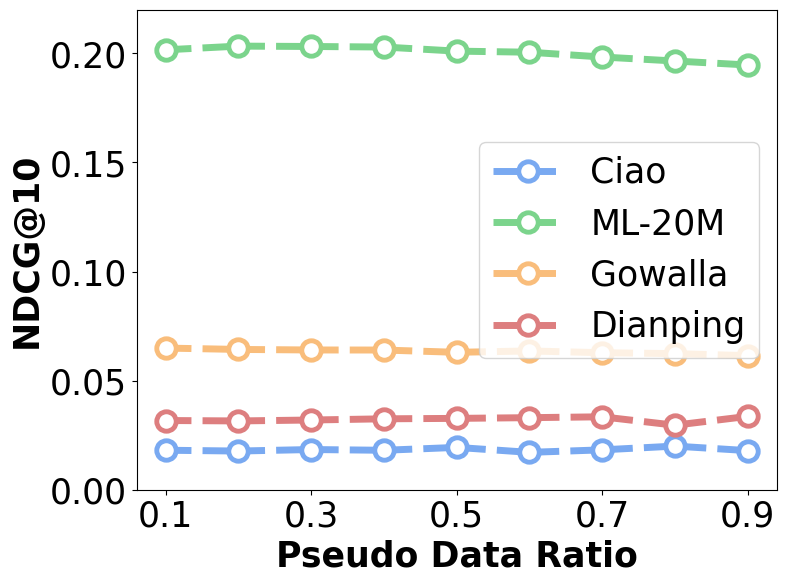}}
\hspace{0.01\columnwidth}
\subfloat[Lightweight update]{\label{fig:dianping-pseudo-ratio-recall}\includegraphics[width=0.495\columnwidth]{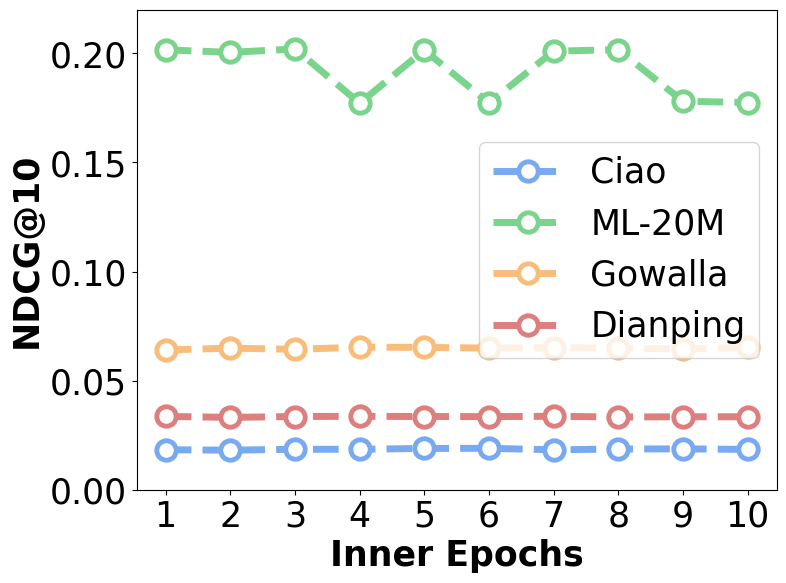}}

% {figures/experiments/pseudo-data-study/Dianping-pseudo-data-ratio-recall.png}}

% \subfigure[Dianping]{\label{fig:dianping-apseudo-ratio}\includegraphics[width=0.4526\columnwidth]{figures/experiments/pseudo-data-study/Dianping-pseudo-data-ratio.png}}
\vskip -0.1in
\caption{Various analysis.}
\label{fig:various-parameter-study}
\vskip -0.1in
\end{figure*}

%% file: sections/sub-sections/experiments/fig-visualization.tex
\begin{figure}[t]
\vskip -0.15in
\centering
\subfloat[Random]{\label{fig:random-visualization}\includegraphics[width=0.333\columnwidth]{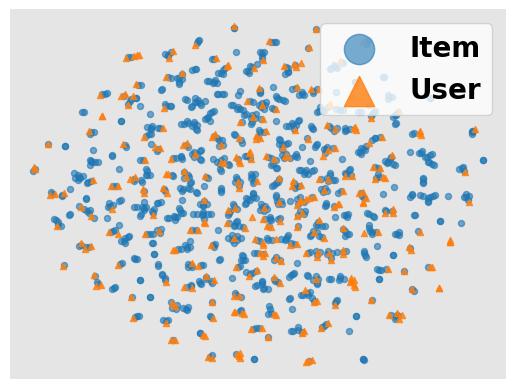}}
\subfloat[\textit{DConRec}]{\label{fig:DCondRec-visualization}\includegraphics[width=0.333\columnwidth]{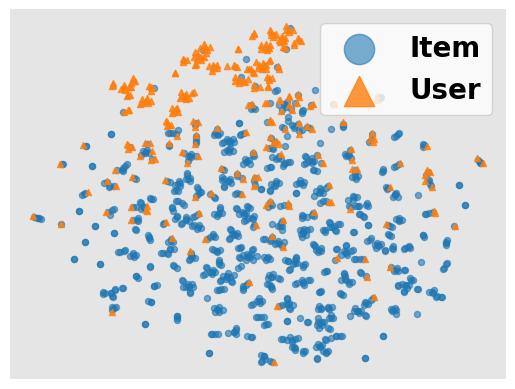}}
% \vskip -0.05in
\subfloat[Full Dataset]{\label{Full-dataset-visualization}\includegraphics[width=0.333\columnwidth]{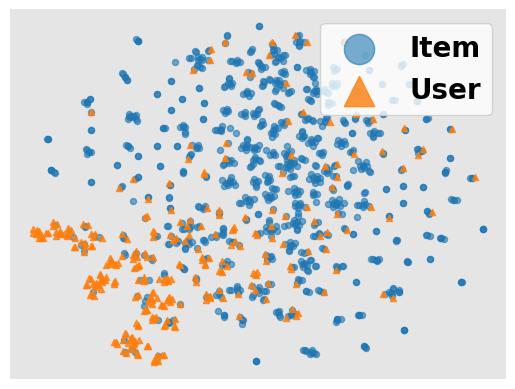}}
\vskip -0.1in
\caption{T-SNE visualization of embeddings learned by MF.}
\label{fig:t-sne-visualization}
\vskip -0.2in
\end{figure}

%% file: sections/relatedWork.tex
\section{Related Work}
\label{sec:related-work}
% \textbf{Recommendation.} \red{recommendation->ID-based recommendation->sigir23's literature review}

% \noindent\textbf{Coreset Selection.} \red{**revise}The conventional paradigm of selection-based methods, (i.e., coreset selection~\cite{wsdm22SVP-data-gene,PCoreset-icml2022}), involves the strategic selection of a subset of data points. The primary objective is to ensure that the selected subset retains the same level of informativeness as the full dataset. Within this domain, three prevalent types of methods have been identified~\cite{wsdm22SVP-data-gene}.
% The first is submodular approaches, which aim to optimize a particular function. The optimized function serves as a measure of utility for the data subset~\cite{wacv2019Coreset,wenjing23arXiv,ieeeComp23shengcai}. In contrast, some works model coreset selection as a bi-level optimization problem~\cite{nips2020BilevelCoreset,ICASSP2021Bilevel,msu23LLMGraph,zgh23LLMRobust}, directly constructing the coreset for specific downstream tasks. Another technique is selection-via-proxy, wherein a base model is utilized as a proxy to identify importance tags to individual data points~\cite{wsdm22SVP-data-gene,iclr2020SVPOriginal,jiajun22icml,liu2021effective}.
% However, being heuristic-based, they lack a guarantee of optimal solutions for recommendation. Moreover, coreset selection assumes that dataset information is concentrated on a few samples, which may not hold true in scenarios like recommendation.\\
\noindent\textbf{Recommender Systems} 
 A prevalent paradigm of the recommendation always decomposes the interaction data into embedding vectors of users and items. One of the most popular techniques is collaborative filtering~\cite{fan2019graphRec,wu-et-al:DcRec_cikm22,jiajun2023iclr,tkde2023GDCF4SoRec,tkde2023MMDisCF,wang2023braveTpami}. It is based on the assumption that users behaving similarly are likely to share similar preference~\cite{TKDE2023lewuSurvey}. For instance, the pioneering study (matrix factorization)~\cite{icdm2010RendleMF,rendle-et-al:bpr2009} aims to learn representations for each user and item, performing inner product between them to predict the preference. Later on, deep neural networks~\cite{he-et-al:NeuMF17he,jiajun22icml,ruihe23ITETCI,pi2-2020jiajun,msu23LLMGraph,zgh23LLMRobust,exgc2024www-junfeng} brought promising application in recommendation tasks. For example, NeuMF~\cite{he-et-al:NeuMF17he} is proposed to replace the inner product of MF model with non-linear neural networks to predict the preference of users. Then, NGCF~\cite{sigir2019NGCF} models the user-item interactions as a graph and adopt the message passing among the nodes to learn representations. Further, LightGCN~\cite{he-et-al:lightgcn} proposes to eliminate the feature transformation and non-linear activation of NGCF. 
 Inspired by the success of self-supervised learning (SSL), the XSimGCL~\cite{XSimGCL2022TKDE} devises a SSL-enhanced method for recommendation, achieving promising performance. \\
 % Those methods could be roughly divided into two categories~\cite{IDvsGenerative2023}: non-sequential models (NSM) and sequential recommendation models (SRM). NSM further includes various recall (i.e., DSSM, Youtube DNN~\cite{recSys2016YoutubeDNN}), CTR models (i.e., DeepFM~\cite{RuimingTang2017DeepFM}, Deep Crossing~\cite{KDD2016Cross}) and other types (i.e., graph-based models~\cite{he-et-al:lightgcn,fan2022graphTrend,fan2019graphRec}). On the other hand, a typical paradigm of SRM is to predict the probability of the next interaction based on a sequence of user-item interactions. The widely-known sequential recommendation models include GRU4Rec~\cite{GRU4Rec}, NextItNet~\cite{sigir2021nextItemNet,wsdm2019NetitNet}, SR-GNN~\cite{aaai2019SR-GNN}, SASRec~\cite{ICDM2018SASRec} and BERT4Rec~\cite{CIKM2019Bert4Rec} with RNN, CNN, GNN, Transformer and BERT as the backbone, respectively. \\
% Graph Trend Filtering Networks for Recommendation
% \red{related work should also be revised and add a section for recommendation}
\textbf{Dataset Condensation}. The goal of dataset condensation is to synthesize a small set of synthetic samples, whose training performance on deep neural networks can be comparable with the original one. There are two types of condensation methods~\cite{pmlr2023ScaleUpDD,junfeng2023nips}: matching-based approaches and kernel-based approaches. 
The first line of research is kernel-based approaches. Some works \cite{infiniteRec22Nips,iclr2021KRR} reduce the bi-level optimization into a single-level one via kernel ridge regression (NTK). Distill-CF~\cite{infiniteRec22Nips} utilizes NTK to model recommendation data and synthesizes fake users via reconstruction. 
%Our proposed method is different from those methods, which adopts one-step update strategy for the inner optimization, inherently reducing the total number of training iterations. 
As for another line, gradient matchings between the model trained on small-condensed dataset and the large-real dataset is proposed by~\cite{zhaoICLR2021DC,icml2021DC-Siamese} to generate synthetic datasets. To avoid high computation of matching gradients, \cite{DZhaoBo2023WACVDistribution} proposes feature matching. Some researchers investigate condensing graphs~\cite{doscond-kdd2022,New2023GraphCondense,graph2022ReceptiveFieldMatching,gcSurvey2024IJCAI,GC-bench2024arXiv,GCSurveyUQ2024arXiv,PartitionGC2024arXiv}. To make the condensation more flexible and adaptable to different architectures of models, some studies~\cite{generative2022DD,cazenavette2023GenerativeDD} propose methods that are based on generative model. GCM~\cite{ctrDD2023RecSys} proposes an efficient condensation method for data in CTR predictions. %Furthermore, some works explore to condense the large scale datasets~\cite{pmlr2023ScaleUpDD,SqueezeRR2023InSubmission}. 
Since they involve expensive and verbose gradient flow, we propose the forward lightweight policy gradient estimator for data updates, which is significantly faster. Besides, to protect user's potential preference, we synthesize data based on pre-augmented data instead of original dataset.% to preserve the potential interests of users.
% \red{add a subsection for recommendation}
% \red{Our method is blablabla}\\
% Scaling Up Dataset Distillation to ImageNet-1K with Constant Memory\\
% Condensing Graphs via One-Step Gradient Matching\\
% Slimmable Dataset Condensation
% \subsection{Data Sampling \red{Coreset Selection}}

%% file: sections/conclusion.tex
\section{Conclusion}
\label{sec:conclusion}
% In this paper, we investigate the topic of condensing recommendation dataset, which is rarely explored before. To be specific, we identify several key challenges in the process of condensing recommendation dataset: high computational burden of condensation regarding of billions of user-item interactions, possible amplification of long-tailed distribution and bad effects of false negative samples. 
% To addresses these issues, we propose a novel framework \textit{\ournameAbbr} that adopts one-step policy gradient estimator to reduce computational burden, introduces compensation for long-tailed users and identifies false negative samples via a simple proxy model. We further verify the effectiveness of the proposed framework empirically and analysis the convergence property from a theoretical perspective. \red{Notably, we performance is blablablalbalbala}
% \red{Conclusion should also be revised}
%we identify the key challenges in dataset condensation for recommendation. They are the high computational burden, the potential amplification of long-tailed distribution patterns, and the detrimental effects of false negative items. To tackle these intricate issues, 
In this paper, we proposed a novel condensation framework for recommendation data, called \ournameAbbr, aiming to condensing interaction set. It can efficiently generate discrete interactions based on pre-augmented data via lightweight policy gradient estimation. Our empirical investigations verify the effectiveness and efficiency of the proposed framework. Notably, we reduce the dataset size by 75\% while approximating over 98\% of the original performance on Dianping and over 90\% on other datasets. Besides, our method is significantly faster than the prevalent gradient matching (e.g., $8\times$ on Ciao). Moreover, we examine its convergence.

%% file: sections/acknowledge.tex
\section{achknowledgements}
% \begin{acks}
This work was supported in part by the National Key Research
and Development Program of China under Grant 2022YFA1004102, and in
part by the Guangdong Major Project of Basic and Applied Basic Research
under Grant 2023B0303000010. Also, the research described in this paper has been partly supported by the National Natural Science Foundation of China (project no. 62102335), General Research Funds from the Hong Kong Research Grants Council (project no. PolyU 15200021, 15207322, and 15200023), internal research funds from The Hong Kong Polytechnic University (project no. P0036200, P0042693, P0048625, P0048752, and P0051361), and SHTM Interdisciplinary Large Grant (project no. P0043302). 
% \end{acks}

%% file: sections/author-bio.tex
{\begin{IEEEbiography}[{\includegraphics[width=1in,height=1.25in,clip,keepaspectratio]{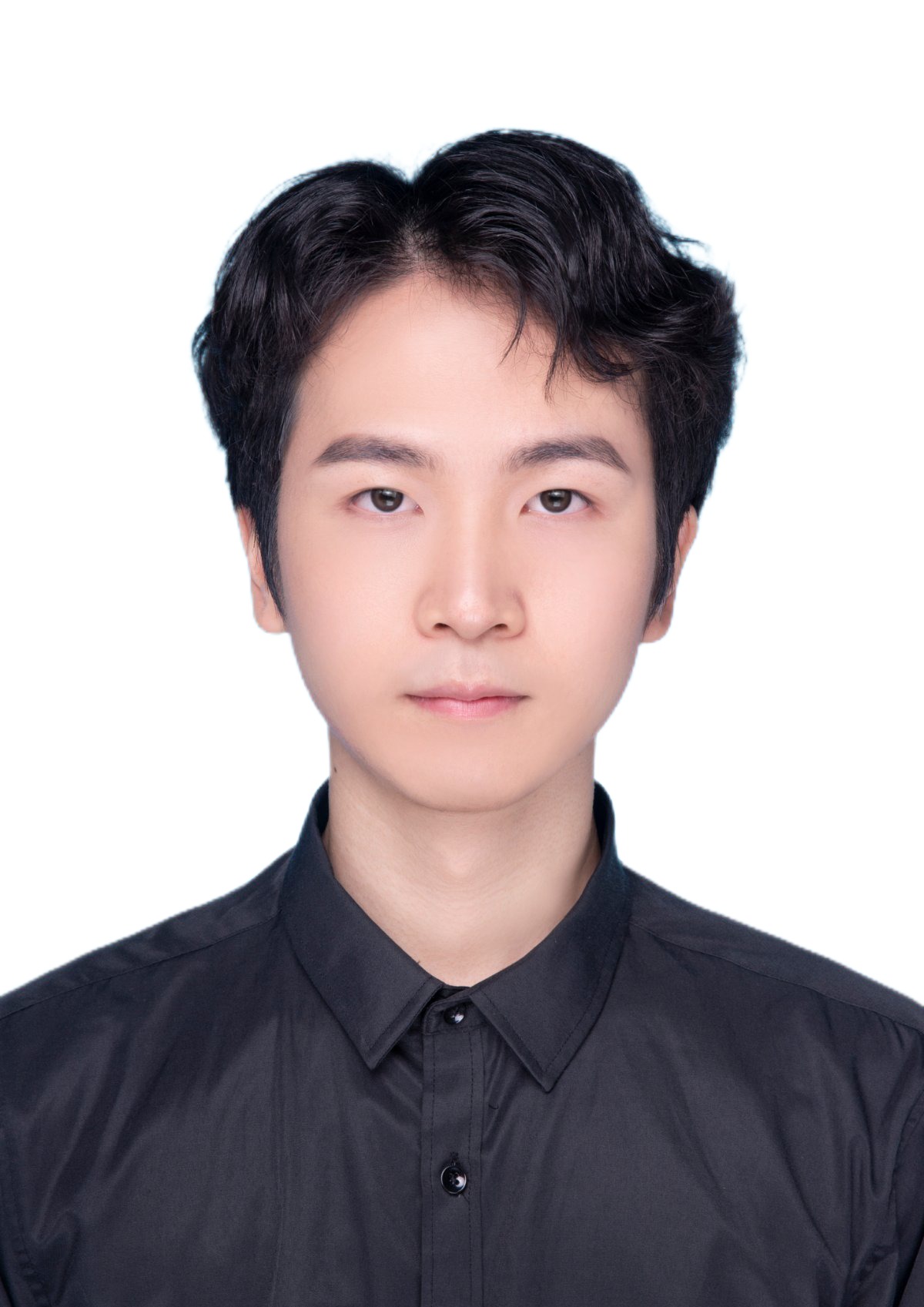}}]{Jiahao Wu} is currently a four-year Ph.D. candidate at The Hong Kong Polytechnic University (PolyU), Hong Kong, and Southern University of Science and Technology (SUSTech), Shenzhen, China. Previously, he earned his B.Eng. degree from the University of Science and Technology of China (USTC), Hefei, China, in 2020. His research interests include recommender systems and dataset condensation. He has innovative works in top-tier conferences. Please find more information at \url{https://jiahaowu.github.io/}.

\end{IEEEbiography}

\vspace{-24pt}
{\begin{IEEEbiography}[{\includegraphics[width=1in,height=1.25in,clip,keepaspectratio]{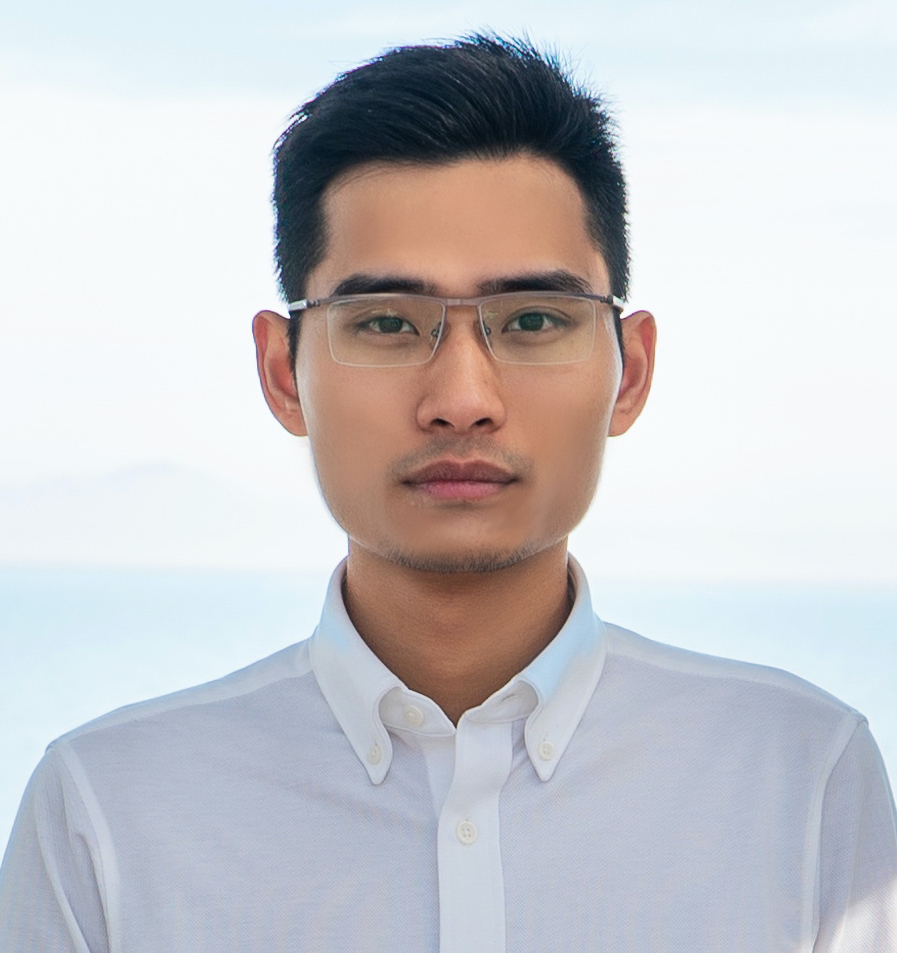}}]{Wenqi Fan} is currently an assistant professor of the Department of Computing at The Hong Kong Polytechnic University (PolyU), Hong Kong. He received his Ph.D. degree from the City University of Hong Kong (CityU), Hong Kong, in 2020. His research interests are in the broad areas of machine learning and data mining, with a particular focus on Recommender Systems, and Large Language Models. He has published innovative papers in top-tier journals and conferences. He serves as a top-tier conference (senior) PC member, session chair, and journal reviewer. More information about him can be found at \url{https://wenqifan03.github.io}.

\end{IEEEbiography}

\vspace{-24pt}

{\begin{IEEEbiography}[{\includegraphics[width=1in,height=1.25in,clip,keepaspectratio]{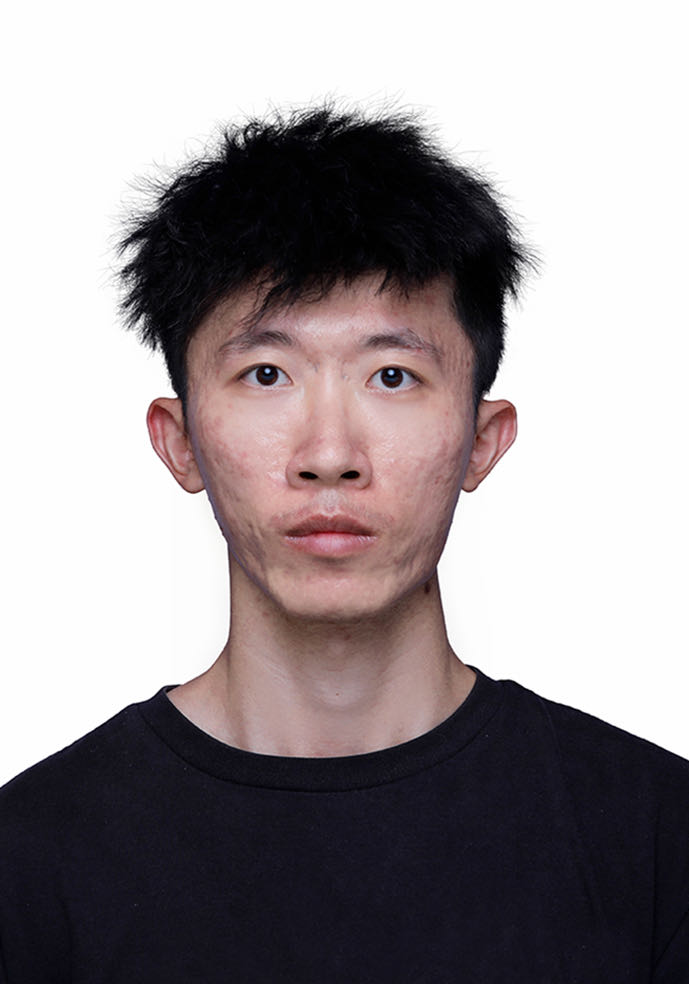}}]{Jingfan Chen} is currently a second-year Ph.D. candidate at The Hong Kong Polytechnic University (PolyU), Hong Kong. Previously, he earned his B.Eng. degree from Shandong University (SDU), Jinan, China, in 2019 and his M.Eng. degree from Nanjing University (NJU), Nanjing, China, in 2022. His research interests include recommender systems, trustworthy AI, and multimodal learning. He has innovative works in top-tier conferences. Please find more information at \url{https://cjfcsjt.github.io/}.

\end{IEEEbiography}

\vspace{-24pt}

{\begin{IEEEbiography}[{\includegraphics[width=1in,height=1.25in,clip,keepaspectratio]{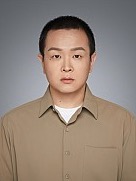}}]{Shengcai Liu} is currently a tenure-track assistant professor with the Department of Computer Science and Engineering, Southern University of Science and Technology (SUSTech), Shenzhen, China. He received Ph.D. degree in computer science and technology from the University of Science and Technology of China (USTC), Hefei, China, in 2020, respectively. His major research interests include learning to optimize and combinatorial optimization. He has authored or co-authored more than 25 papers in top-tier refereed international conferences and journals. More information about him can be found at \url{https://senshinel.github.io/}.

\end{IEEEbiography}

\vspace{-24pt}

{\begin{IEEEbiography}[{\includegraphics[width=1in,height=1.25in,clip,keepaspectratio]{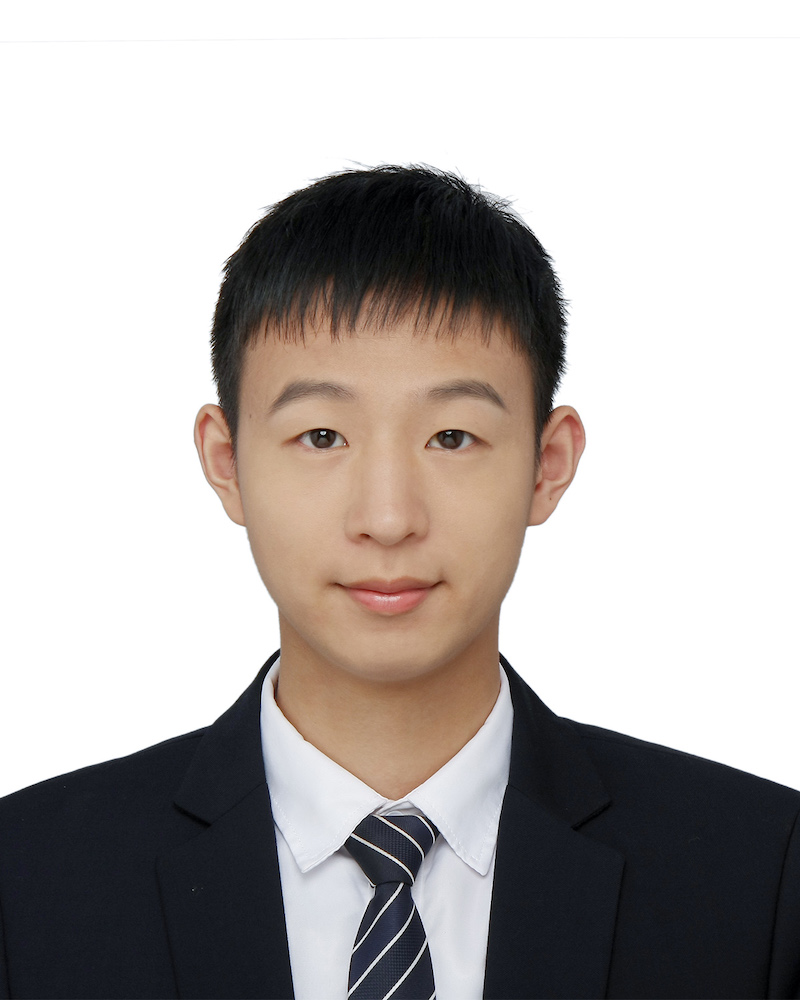}}]{Qijiong Liu} is currently a third-year Ph.D. candidate at The Hong Kong Polytechnic University (PolyU). Previously, he earned his B.Eng. and M.Eng. degrees from Zhejiang University (ZJU), Hangzhou, China, in 2018 and 2021. His research interests include recommender systems, generative retrieval, and multimodal models. He has innovative works in top-tier conferences. Please find more information at \url{https://liu.qijiong.work/resume}.

\end{IEEEbiography}

\vspace{-24pt}

{\begin{IEEEbiography}[{\includegraphics[width=1in,height=1.25in,clip,keepaspectratio]{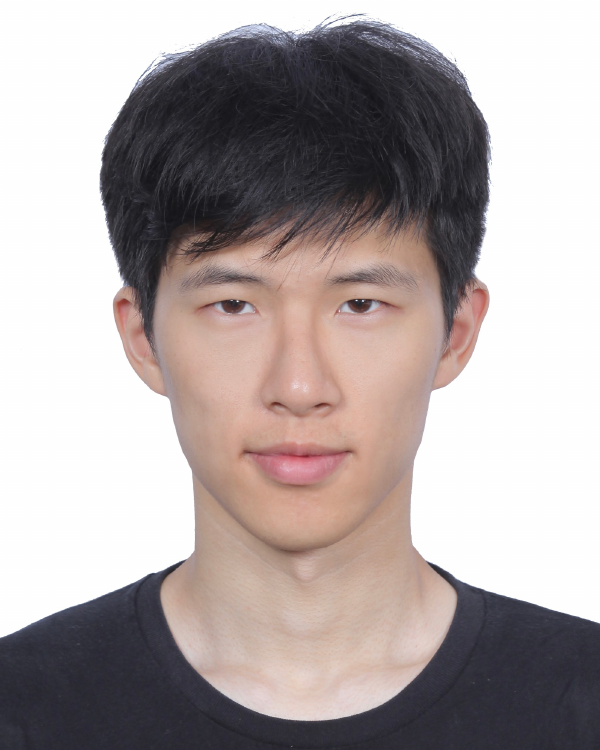}}]{Rui He} is currently a research engineer at BYD Auto. Previously, he obtained a BEng degree from the Southern University of Science and Technology (SUSTech), Shenzhen, China, in 2018 and a Ph.D. degree from the University of Birmingham (UoB) in 2024. His research interest focuses on active learning. He has innovative works in top conferences and journals. Please find more information at \url{https://scholar.google.com/citations?user=9do30b8AAAAJ&hl=en}.

\end{IEEEbiography}

\vspace{-24pt}

{\begin{IEEEbiography}[{\includegraphics[width=1in,height=1.25in,clip,keepaspectratio]{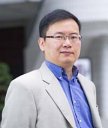}}]{Qing Li} is currently a Chair Professor (Data Science) and the Head of the Department of Computing, at the Hong Kong Polytechnic University (PolyU). He received a Ph.D. degree from the University of Southern California (USC), Los Angeles, USA. His research interests include object modeling, multimedia databases, social media, and recommender systems. He is a Fellow of IEEE and IET and a member of ACM SIGMOD and IEEE Technical Committee on Data Engineering. He has been actively involved in the research community by serving as an associate editor and reviewer for technical journals and as an organizer/co-organizer of numerous international conferences. He is the chairperson of the Hong Kong Web Society, and also served/is serving as an executive committee (EXCO) member of the IEEE-Hong Kong Computer Chapter and ACM Hong Kong Chapter. In addition, he serves as a councilor of the Database Society of Chinese Computer Federation (CCF), a member of the Big Data Expert Committee of CCF, and is a Steering Committee member of DASFAA, ER, ICWL, UMEDIA, and WISE Society. Please find more information at \url{https://www4.comp.polyu.edu.hk/~csqli/}.

\end{IEEEbiography}

\vspace{-24pt}

{\begin{IEEEbiography}[{\includegraphics[width=1in,height=1.25in,clip,keepaspectratio]{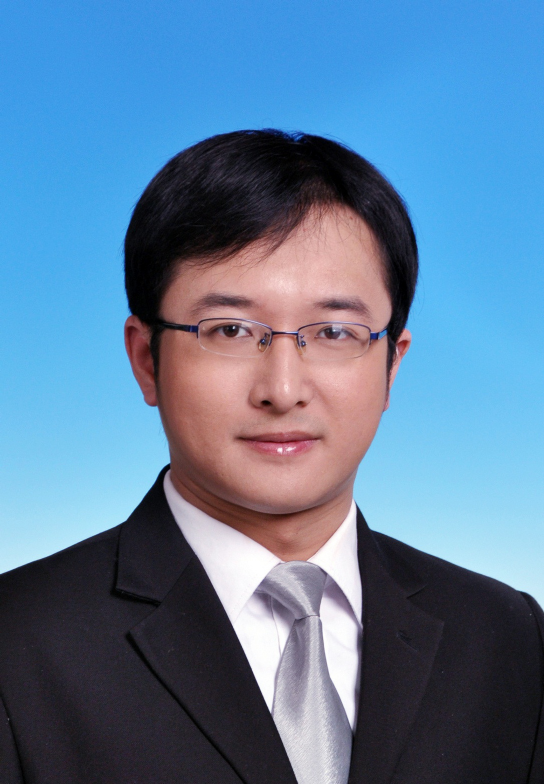}}]{Ke Tang} is currently a professor at the Department of Computer Science and Engineering, at Southern University of Science and Technology (SUSTech), Shenzhen, China. He received a Ph.D. degree in computer science from Nanyang Technological University (NTU), Singapore, in 2007. His research interests mainly include evolutionary computation, machine learning, and their applications. He is a Fellow of IEEE and he was awarded the Newton Advanced Fellowship (Royal Society) and the Changjiang Professorship Ministry of Education (MOE) of China. He was the recipient of the IEEE Computational Intelligence Society Outstanding Early Career Award and the Natural Science Award of MOE of China. He has been actively involved in the research community by serving as an organizer/co-organizer of numerous conferences, an associate editor of the IEEE TRANSACTIONS ON EVOLUTIONARY COMPUTATION, and as a member of editorial boards for a few other journals. Please find more information at \url{https://www.sustech.edu.cn/en/faculties/tangke.html}.

\end{IEEEbiography}

\vspace{-24pt}

%% file: sections/appendix.tex
\appendix
% \section{Dataset Statistics}
% \label{sec:dataset-statistics}
% \input{sections/sub-sections/experiments/tab-datasets}

% Parameter settings.
%  All the experiments are run with
% Pytorch (version 1.12) on Nvidia A30 GPU (Memory: 24GB, Cuda
% version: 11.3).
% \subsection{Implementation Details of Baselines}
% \label{subsec:appendix-doscond-impl-details}
% \input{sections/sub-sections/alg-implementation-doscond}
% \subsection{Dataset Statistics}
% \label{sub-sec: dataset-statistics}

% \section{Algorithm of Projection}
% The projection of $\mathcal{P}_{\mathcal{C}}(\cdot)$ can be calculated as:
\section*{Algorithms}
\input{sections/sub-sections/alg-os-pge}
\input{sections/sub-sections/alg-projection}
\section*{Proof for the Theorem}
\input{sections/sub-sections/append-proof}

%% file: sections/sub-sections/alg-os-pge.tex
\begin{algorithm}
\begin{algorithmic}[1]
\caption{Condensing Rec-Dataset}
\label{algo:training}
\REQUIRE data pool $\mathcal{D}_{p}$, validation dataset $\mathcal{D}_{v}$, a recommender model $\theta$, condensation ratio $r$.
\ENSURE condensed dataset $\mathcal{D}_{s}$.
\STATE Initialize condensation probability $\bs{s}_{u,i}=r$%\bs{1}$.
\STATE Generate compensation dataset $\mathcal{D}_{t}$ from $\mathcal{D}_{p}$.
% \WHILE{Not Convergence}
% \STATE xxxx\\
% \STATE xxxx\\
\FOR{training iteration $t=1,2...T$}
    \STATE Sample matrix $\mathbf{M}$ based on the probability $\mathbf{S}$.\\
    \STATE Randomly initialize the model $\theta_0$.\\
    \STATE Train the model for one epoch:\gray{{//Inner Optimization}}\\
    % \[
    $\quad\quad\quad\quad\quad\theta_1 \leftarrow \theta_0 - \nabla_{\theta_0}\hat{\mathcal{L}}_{rec}(\theta_0;\mathbf{M})$
    % \]
    % \STATE Sample a mini-batch interactions $\mathcal{B}$ from validation dataset $\mathcal{D}_{v}$.
    \STATE Optimize $\mathbf{S}$ via Eq.~\ref{eq:one-step-gradient-s} based on $\mathcal{D}_{v}$:\gray{{//Outer Optimization}}\\
    % \[
    $\quad\quad\mathbf{S}^{t+1} \leftarrow \mathcal{P}_{\mathcal{C}}\left(\mathbf{S}^t-\eta \mathcal{L}_{\mathcal{D}_{v}}%\left(\theta_1(\mathbf{M})\right)
    \nabla_{\mathbf{S}^t} \ln p(\mathbf{M} \mid \mathbf{S}^t)\right)$
    % \]
    \STATE Sample $\mathbf{M}$ based on the probability $\mathbf{S}^{t+1}$.\\
    \STATE Synthesize $\mathcal{D}_s$ based on generation matrix $\mathbf{M}$.
    % \STATE Compensate for long-tailed users: $\mathcal{D}'_{s}\leftarrow\mathcal{D}_s\cup\mathcal{D}_{t}.$
    % \IF{$epoch_{idx}<N_{pre}$}
    % \STATE  xxxx\\
    % \ELSE
    %     \STATE xxxx 
    % \ENDIF
    % \STATE xxxxx \\
    % \STATE  xxxx   / xxxx\\
\ENDFOR
% \ENDWHILE
% \STATE ${z}_u$, ${z}_i=$  CF-Backbone($\mathcal{G}$, ${z}^{(0)}_u$, ${z}^{(0)}_i$); {// Eq.~(\ref{eq:graph_prop})-(\ref{eq:layer_sum})}\\
\STATE Return the condensed dataset $\mathcal{D}_{s}$
\end{algorithmic}
\end{algorithm}
% \vskip -0.2in

%% file: sections/sub-sections/alg-projection.tex
\begin{algorithm}
\begin{algorithmic}[1]
\caption{Algorithm of Projection $\mathcal{P}_{\mathcal{C}}(\cdot)$}
\label{algo:projection}
\REQUIRE a matrix $\mathbf{S}_{in}$, condensation ratio $r$.
\ENSURE projected matrix $\mathbf{S}_{out}$.
% \Procedure{Critical-Sample}{$\mathcal{D}$,$\boldsymbol{\theta}$}
    \STATE Calculate $v_1$ via:\\
    % \[
    $\quad\quad\mathbf{1}^T\left[\min\left(1,\max\left(0,\mathbf{S}_{in}-v_1\mathbf{1}\right)\right)\right]-r|\mathcal{D}|=0.$
    % \]
    \STATE $v_2\leftarrow \max(0,v_1)$.
    \STATE $\mathbf{S}_{out}\leftarrow \min\left(1,\max\left(0,\mathbf{S}_{in}-v_2\mathbf{1}\right)\right)$.
\STATE return $\mathbf{S}_{out}$
\end{algorithmic}
% \vskip -0.2in
\end{algorithm}
% \vskip -0.2in

%% file: sections/sub-sections/append-proof.tex
\section{Proof of the Theorem}
\label{sec:appendix-proof}
% In this section, we provide theoretical proof for the theorem~\ref{thm:convergence} presented in section~\ref{sec:framework}. The proof is inspired and based on~\cite{doscond-kdd2022,icml-Pedregosa16}. 
Before proving the theorem, we introduce two lemmas about the properties of the projection operator $\mathcal{P}_{\mathcal{C}}(\cdot)$~\cite{icml-Pedregosa16}.
% \textbf{Lemmas:}
\begin{lemma}
    \label{lemma:lemma1}
     (firmly nonexpansive operators). Given a compact convex set $\mathcal{C}\subset\mathbb{R}^d$ and let $\mathcal{P}_{\mathcal{C}}(\cdot)$ be the projection operator on $\mathcal{C}$, then for any $\boldsymbol{u} \in \mathbb{R}^d$ and $\boldsymbol{v} \in \mathbb{R}^d$, we have\[
     \left\|\mathcal{P}_{\mathcal{C}}(\boldsymbol{u})-\mathcal{P}_{\mathcal{C}}(\boldsymbol{v})\right\|^2 \leq(\boldsymbol{u}-\boldsymbol{v})^{\top}\left(\mathcal{P}_{\mathcal{C}}(\boldsymbol{u})-\mathcal{P}_{\mathcal{C}}(\boldsymbol{v})\right).
     \]
\end{lemma}
\begin{lemma}
    \label{lemma:lemma2}
    Given a compact convex set $\mathcal{C}\subset\mathbb{R}^d$ and let $\mathcal{P}_{\mathcal{C}}(\cdot)$ be the projection operator on $\mathcal{C}$, then for any $\boldsymbol{c}\in \mathcal{C}$ and $\boldsymbol{u} \in \mathbb{R}^d$, $\boldsymbol{v} \in \mathbb{R}^d$, we have\[
    \left\|\mathcal{P}_{\mathcal{C}}(\boldsymbol{c}+\boldsymbol{u})-\mathcal{P}_{\mathcal{C}}(\boldsymbol{c}+\boldsymbol{v})\right\| \leq\|\boldsymbol{u}-\boldsymbol{v}\|.
    \]
\end{lemma}

\noindent Then, we provide the theorem proof, which is partially motivated by~\cite{doscond-kdd2022,PCoreset-icml2022}.
\allowdisplaybreaks[4]
\begin{proof}
    \label{pf: proof-theorem}
    In the implementation of algorithm~\ref{algo:training}, we calculate the gradient for updating $\mathbf{S}$ based on validation set $\mathcal{D}_{val}$, which is a subset randomly sampled from $\mathcal{D}$. Therefore, in the following, we denote the gradient via $\boldsymbol{g}^t$:\[
\boldsymbol{g}^t=\mathcal{L}_{val}\left(\boldsymbol{\theta}_1(\boldsymbol{m})\right) \nabla_{\mathbf{S}} \ln p\left(\boldsymbol{m} \mid \mathbf{S}^t\right).
    \]
    Consequently, the updating of $\mathbf{S}$ in algorithm~\ref{algo:training} could be re-written as:\[
    \mathbf{S}^{t+1} = \mathcal{P}_{\mathcal{C}}\left(\mathbf{S}^t-\eta \boldsymbol{g}^t\right).
    \]
    Let us denote the validation set based gradient mapping be:\[
    \hat{\mathcal{G}}^t=\frac{1}{\eta}\left(\mathbf{S}^t-\mathcal{P}_{\mathcal{C}}\left(\mathbf{S}^t-\eta \boldsymbol{g}^t\right)\right)=\frac{1}{\eta}\left(\mathbf{S}^t-\mathbf{S}^{t+1}\right),
    \]
    we have: \begin{align*}
    % \scalebox{0.75}{
    % \resizebox{.9\hsize}{!}{
    &\Phi(\mathbf{S}^{t+1})\leq\Phi(\mathbf{S}^t) + \left<\nabla\Phi(\mathbf{S}^t),\mathbf{S}^{t+1}-\mathbf{S}^t\right>+\frac{L}{2}\lVert\mathbf{S}^{t+1}-\mathbf{S}^t\rVert^2\\%\gray{//{L-smoothness}}
    &=\Phi(\mathbf{S}^t) - \eta \left<\nabla\Phi(\mathbf{S}^t),\hat{\mathcal{G}}^t\right>+\frac{L\eta^2}{2}\lVert\hat{\mathcal{G}}^t\rVert^2\\
    &=\Phi(\mathbf{S}^t) - \eta \left<\nabla\Phi(\mathbf{S}^t)-\boldsymbol{g}^t+\boldsymbol{g}^t,\hat{\mathcal{G}}^t\right>+\frac{L\eta^2}{2}\lVert\hat{\mathcal{G}}^t\lVert^2\\
    &=\Phi(\mathbf{S}^t) - \eta \left<\boldsymbol{g}^t,\hat{\mathcal{G}}^t\right>+\frac{L\eta^2}{2}\lVert\hat{\mathcal{G}}^t\rVert^2+\eta\left<\delta^t,\hat{\mathcal{G}}^t\right>\\%\gray{//\delta^t=\boldsymbol{g}^t-\nabla\Phi(\mathbf{S}^t)}
    &\leq\Phi(\mathbf{S}^t) - \eta\lVert\hat{\mathcal{G}}^t\rVert^2+\frac{L\eta^2}{2}\lVert\hat{\mathcal{G}}^t\rVert^2+\eta\left<\delta^t,\hat{\mathcal{G}}^t\right>\gray{//{Lemma~\ref{lemma:lemma1}}}\\
    &=\Phi(\mathbf{S}^t) - (\eta-\frac{L\eta
    ^2}{2})\lVert\hat{\mathcal{G}}^t\rVert^2+\eta\left<\delta^t,\hat{\mathcal{G}}^t\right>\\
    &=\Phi(\mathbf{S}^t) - (\eta-\frac{L\eta
    ^2}{2})\lVert\hat{\mathcal{G}}^t\rVert^2+\eta\left<\delta^t,\mathcal{G}^t\right>+\eta\left<\delta^t,\hat{\mathcal{G}}^t-\mathcal{G}^t\right>\\
    &\leq \Phi(\mathbf{S}^t) - (\eta-\frac{L\eta
    ^2}{2})\lVert\hat{\mathcal{G}}^t\rVert^2+\eta\left<\delta^t,\mathcal{G}^t\right>+\eta\lVert\delta^t\rVert \lVert\hat{\mathcal{G}}^t-\mathcal{G}^t\rVert\\
    &\leq \Phi(\mathbf{S}^t) - (\eta-\frac{L\eta
    ^2}{2})\lVert\hat{\mathcal{G}}^t\rVert^2+\eta\left<\delta^t,\mathcal{G}^t\right>+\eta\lVert\delta^t\rVert^2.\gray{//{Lemma~\ref{lemma:lemma2}}}
    \end{align*}
    $\Rightarrow$ %\[
    $(\eta-\frac{L\eta
    ^2}{2})\lVert\hat{\mathcal{G}}^t\rVert^2\leq\Phi(\mathbf{S}^t) - \Phi(\mathbf{S}^{t+1})+\eta\left<\delta^t,\mathcal{G}^t\right>+\eta\lVert\delta^t\rVert^2.$\\
    %\]
    Then, we get:
        \begin{align*}
            \label{eq:proof-theorem-eq1}
            &\sum_{t=1}^{T}(\eta-\frac{L\eta
    ^2}{2})\lVert\hat{\mathcal{G}}^t\rVert^2 \\&\leq\Phi(\mathbf{S}^1) - \Phi(\mathbf{S}^{T+1})+\eta\sum_{t=1}^{T}\left(\left<\delta^t,\mathcal{G}^t\right>+\eta\lVert\delta^t\rVert^2\right).
        \end{align*}
    Since,
    \begin{equation}
        \label{eq:proof-theorem-eq2}
        \mathbb{E}\left<\delta^t,\mathcal{G}^t\right>=\mathbb{E}_{\mathbf{S}^t}\mathbb{E}_{\cdot|\mathbf{S}^t}\left(\left<\boldsymbol{g}^t-\nabla\Phi(\mathbf{S}^t)),\mathcal{G}^t\right>|\mathbf{S}^t\right)=0,
    \end{equation}
    and based on the assumption described as following:
    \begin{equation}
        \label{eq:proof-theorem-eq3}
        \mathbb{E}\lVert\delta^t\rVert^2=\mathbb{E}\lVert\boldsymbol{g}^t-\nabla\Phi(\mathbf{S}^t)\rVert^2\leq \sigma^2.
    \end{equation}
    we can get follows by by taking Eq.~\ref{eq:proof-theorem-eq2} and Eq.~\ref{eq:proof-theorem-eq3} back into Eq.~\ref{eq:proof-theorem-eq3}:\begin{equation}
        \label{eq:proof-theorem-eq4}
        \frac{1}{T}\sum_{t=1}^{T}\mathbb{E}\lVert\hat{\mathcal{G}^t}\rVert^2\leq\frac{\Phi(\mathbf{S}^1)-\Phi^*}{(1-L\eta/2)T}+\frac{\sigma^2}{1-L\eta/2}.
    \end{equation}   
    Since we have get:\begin{align}
        \label{eq:proof-theorem-eq5}
        % \begin{aligned}
            \mathbb{E}\lVert\mathcal{G}^t\rVert^2&\leq 2\mathbb{E}\lVert\hat{\mathcal{G}^t}\rVert^2 + 2\mathbb{E}\lVert\boldsymbol{g}^t-\nabla\Phi(\mathbf{S}^t)\rVert^2 \leq2\mathbb{E}\lVert\hat{\mathcal{G}^t}\rVert^2 + 2\sigma^2,
        % \end{aligned}
    \end{align}
    we can obtain follows by taking Eq.~\ref{eq:proof-theorem-eq4} into Eq.~\ref{eq:proof-theorem-eq5} and letting $T\rightarrow \infty$ to have:\begin{equation*}
        \begin{aligned}
            \frac{1}{T}\sum_{t=1}^{T}\mathbb{E}\lVert\mathcal{G}^t\rVert^2&\leq\frac{2}{1-L\eta/2}\left(\frac{\Phi(\mathbf{S}^1)-\Phi^*}{T}+(2-L\eta/2)\sigma^2\right)\\
            &\rightarrow \frac{8-2 L \eta}{2-L \eta} \sigma^2.
        \end{aligned}
    \end{equation*}
\end{proof}